\documentclass[twocolumn,showpacs,preprintnumbers,amsmath,amssymb,prb]{revtex4}
\usepackage{graphicx}% Include figure files
\usepackage{dcolumn}% Align table columns on decimal point
\usepackage[tight]{subfigure}
\usepackage{amsmath}
\usepackage{verbatim}
\usepackage[usenames,dvipsnames]{color}
\usepackage{bm} % bold greek math
\usepackage{bbm}
\newcommand{\assign}{:=}

\newcommand{\substacktwo}[2]{
  \substack{ \mathrm{#1} \\ \mathrm{#2}}
}
\newcommand{\Figure}[2]{
  \begin{figure}[ht]
    \includegraphics[width=0.9\linewidth]{#1}
    \caption{#2}
  \end{figure}
}
\expandafter\newcommand\csname Figure2\endcsname[3]{
\begin{figure}[ht]
  \includegraphics[width=0.9\linewidth]{#1}
  \\
  \includegraphics[width=0.9\linewidth]{#2}
  \caption{#3}
\end{figure}
}

\newcommand{\wideeq}[1]{\begin{widetext}
    #1
  \end{widetext}}

\newcommand{\Fig}[1]{Fig.~\ref{#1}}
\newcommand{\fig}[1]{\ref{#1}}
\newcommand{\Eq}[1]{Eq.~(\ref{#1})}
\newcommand{\eq}[1]{(\ref{#1})}
\newcommand{\Sec}[1]{Sec.~\ref{#1}}

\newcommand{\App}[1]{Appendix~\ref{#1}}

\newcommand{\Cite}[1]{Ref.~\onlinecite{#1}}
\newcommand{\Cites}[1]{Refs.~\onlinecite{#1}}

\newcommand{\todo}[1]{ \textbf{\textcolor{Bittersweet}{TODO: #1}} }
\newcommand{\hide}[1]{ \textbf{\textcolor{Gray}{#1}} }

\renewcommand{\todo}[1]{}
\renewcommand{\hide}[1]{}
\newcommand{\op}[1]{\hat{#1}}
\renewcommand{\vec}[1]{\mathbf{#1}}
\newcommand{\vecg}[1]{{\bm #1}}

\newcommand{\bra}[1]{\langle #1|}
\newcommand{\ket}[1]{|#1 \rangle}
\newcommand{\brkt}[1]{\langle #1 \rangle}
\newcommand{\Tr}{\text{Tr}}
\newcommand{\tr}[1]{\underset{#1}{\text{Tr}}}

\begin{document}
\title{
  Coherent Backaction of Quantum Dot Detectors: Qubit Isospin Precession
}

\author{M. Hell$^{(1,2)}$}
\author{M. R. Wegewijs$^{(1,2,3)}$}
\author{D. P. DiVincenzo$^{(1,2,4)}$}
\affiliation{
  (1) Peter Gr{\"u}nberg Institut,
      Forschungszentrum J{\"u}lich, 52425 J{\"u}lich,  Germany
  \\
  (2) JARA- Fundamentals of Future Information Technology
  \\
  (3) Institute for Theory of Statistical Physics,
      RWTH Aachen, 52056 Aachen,  Germany
  \\
  (4) Institute for Quantum Information,
      RWTH Aachen, 52056 Aachen,  Germany
}
\date{\today}

\begin{abstract}
  A sensitive technique for the readout of the state of a qubit is based on
  the measurement of the conductance through a proximal sensor quantum dot
  (SQD). Here, we theoretically study the {\emph{coherent}} backaction of such
  a measurement on a coupled SQD-charge-qubit system. We derive Markovian
  kinetic equations for the ensemble-averaged state of the SQD-qubit system,
  expressed in the coupled dynamics of two charge-state occupations of
  the SQD and two qubit isospin vectors, one for each SQD charge
  state. We find that aside from introducing dissipation, the detection also
  renormalizes the coherent evolution of the SQD-qubit system. Basically, if
  the electron on the detector has time to probe the qubit, then it also has
  time to fluctuate and thereby renormalize the system parameters. In
  particular, this induces {\emph{torques}} on the qubit isospins, similar to
  the spin torque generated by the spintronic exchange field in noncollinear
  spin-valve structures. Secondly, we show that for a consistent
  description of the detection, one must also include the renormalization
  effects in the \emph{next-to-leading} order in the electron
  tunneling rates, especially at the point of maximal sensitivity of the
  detector. Although we focus on a charge-qubit model, our findings are
  generic for qubit readout schemes that are based on spin-to-charge
  conversion using a quantum-dot detector. Furthermore, our study of
  the stationary current through the SQD, a test measurement verifying
  that the qubit couples to the detector current, already reveals
  various significant effects of the isospin torques on the
  qubit. Our kinetic equations provide a starting point for further
  studies of the time evolution in charge-based qubit readout. Finally,
  we provide a rigorous sum rule that constrains such approximate
  descriptions of the qubit isospin dynamics and show that it is obeyed by
  our kinetic equations.
\end{abstract}

 \maketitle

\section{Introduction}\label{sec:intro}

Quantum computation demands the readout of the state of a quantum bit (qubit)
with high fidelity. In principle, this can be realized exclusively using
all-electric components by spin singlet-triplet qubits {\cite{Levy02,Petta05}}
manipulated by gate voltages and read out by spin-to-charge conversion. This
technique utilizes a capacitive coupling of the qubit to a nearby quantum
point contact (QPC) {\cite{Barthel09,Reilly07}} or a sensor quantum dot
(SQD) {\cite{Barthel10}}. The latter is advantageous due to its higher
sensitivity, resulting in larger signal-to-noise ratio {\cite{Barthel10}}.
Charge sensing by radio-frequency single-electron transistors (RF-SETs), first
introduced in {\Cite{Schoelkopf98}} as a static electrometer, moreover allows
real-time observation \ of electron tunneling events,{\cite{Lu03a}} which can
be applied to \ measure ultra-small currents,{\cite{Fujisawa04}} to test
fluctuation relations in electronic systems{\cite{Saira12}} or as which-path
detectors in an Aharonov-Bohm ring.{\cite{Buks98}} However, in all these
setups the measured system suffers from dephasing by the environment, which
leads to a cumulative error that is eventually beyond the reach of quantum
error-correction schemes. Yet, such {\emph{dissipative}} environmental
backaction effects can also be controlled, as for example demonstrated by the
destruction of Aharonov-Bohm oscillations.{\cite{Moldoveanu07,Buks98}} This
may even offer new prospects for qubit control, e.g., by mediating effective
interactions between qubits that can be implemented to engender
entanglement.{\cite{Braun02,Contreras08,Vorrath03,Lambert07}} In addition,
quantum memories may be realized by engineering quantum states through
dissipation {\cite{Pastawski11}}.

Similar dissipative environmental effects are also well-known from
nonequilibrium transport through quantum dots (QDs). However, when a QD is
embedded into a spintronic device with ferromagnetic electrodes, dissipative
effects are not the only way in which it is influenced by the environment,
even in leading order in the coupling: spin-dependent scattering and Coulomb
interaction lead to the generation of a spin torque. This torque derives from
{\emph{coherent}} processes that renormalize the QD energy levels
{\cite{Braun04set,Koenig03,Donarini10}}, resulting in an \ effective magnetic
field, known as the {\emph{spintronic}} exchange
field.{\cite{Braun04set,Koenig03}} The dynamical consequence of the torque is
a precession of the average spin vector on the QD,{\cite{Koenig03}} in
addition to the shrinking of the spin magnitude, which is a pure dissipative
effect. Renormalization effects have not only been discussed for spintronic
devices, but also for STM setups {\cite{Sobczyk12}} and superconducting
nanostructures.{\cite{Governale08,Futterer13}} In the latter case effective
magnetic fields act on an {\emph{isospin}} that describes proximity-induced
coherences between different charge states on the
QD.{\cite{Governale08,Futterer13}} Moreover, environmentally-induced torques
are not only limited to fermionic systems; they have also been discussed for
optical activity.{\cite{Harris82}} In the context of quantum dot readout, they
have been considered for QPCs.{\cite{Stodolsky99}}

It is therefore natural to ask whether similar coherent effects arise when a
qubit is measured by a SQD since any type of readout requires
\emph{interaction} of the system with its detector, which may lead to
renormalization effects. This is the main focus of this paper: we derive and
discuss kinetic equations for the reduced density matrix of the composite
system of SQD and qubit by integrating out the lead degrees of freedom and
employing a Markov approximation. Related previous works have studied, e.g.,
decoherence effects for electrostatic qubits {\cite{Gurvitz05,Gurvitz08}}, or,
Josephson junction qubits {\cite{Shnirman98,Makhlin01a}} focusing on
time-dependent phenomena. In our study, we address the {\emph{continuous}}
measurement limit, \ in which the qubit level splitting $\Omega$ and the
SQD-qubit coupling $\lambda$ are small compared to the single-electron
tunneling (SET) rates $\Gamma$ through the SQD: $\lambda, \Omega \ll \Gamma$.
In this limit, each electron ``sees'' a snapshot of the qubit state as the
qubit evolution is negligible during the interaction time with an electron on
the detector. Our results are not only limited to weak measurements ($\lambda
\ll \Omega$) but are also valid for $\lambda \lesssim \Omega \ll \Gamma$.

We extend previous works in the following four aspects:

(i) We include {\emph{level renormalization}} effects of the qubit
\emph{plus} the detector in the kinetic equations affecting the
energy-nondiagonal part of their density matrix. These effects correspond
physically to \emph{isospin torque} terms that couple the SQD and qubit
dynamics. These torques arise due to the readout processes and cannot be
avoided: they incorporate a term that scales in the same way as the readout
terms $\sim \Gamma \lambda \text{/} T$, where $T$ is temperature. Moreover,
the renormalization of the qubit splitting $\Omega$ leads to additional
induced torque terms $\sim \Gamma \Omega \text{/} T$, expressing the fact that
the charge fluctuations $\sim \Gamma$ are sensitive to all internal energy
scales ($\Omega, \lambda$) of the SQD-qubit system.

(ii) Our kinetic equations also necessarily comprise \emph{next-to-leading
order} corrections to the tunneling $\sim \Gamma^2 \text{/} T$ affecting also
the diagonal part of the density matrix. Generally, these are expected to be
important since maximal sensitivity of the SQD to the qubit state is achieved
by tuning to the flank of the SET peak. In this regime of crossover to Coulomb
blockade, cotunneling broadening and level renormalization effects may compete
with SET processes $\sim \Gamma$. Indeed, in actual readout experiments on
singlet-triplet qubits {\cite{Barthel10}} $\Gamma \sim T$. Moreover, the
inclusion of this renormalization of the SQD tunnel rates is even a mandatory
step since the weaker isospin torque effects $\sim \Gamma \lambda / T, \Gamma
\Omega \text{/} T$ must be included to consistently describe detection at all.

The importance of such an interplay between energy-diagonal and nondiagonal
density matrix parts and higher-order tunnel processes was noted earlier in
Refs. {\cite{Leijnse08a}}, and {\cite{Koller10}}. As in that study, we find
that the failure to account for this leads to severe problems with the
positivity of the density operator in the Markovian
approximation. In standard Born-Markov approaches{\cite{Breuer}} used
to study decoherence effects,{\cite{Chirolli08}} positivity is usually
enforced by a secular approximation.{\cite{Breuer}} As explained in more
detail in {\App{sec:secular}}, a secular approximation is not applicable in
our case because the tunneling rate is not assumed to be small ($\Gamma \gg
\Omega, \lambda$). Despite this, the positivity of the density matrix is \
ensured when consistently including corrections $\sim \Gamma^2 \text{/} T$, as
shown in {\App{sec:sumRule}}.

(iii) In extension to Refs.{~\onlinecite{Gurvitz05,Gurvitz08,Shnirman98,Makhlin01a}}, we
include the electron spin degree of the SQD into our study. This has several
consequences, most notably, for the qubit-dependent part of the current
through the SQD: this current does {\emph{not}} directly measure the qubit
isospin, but charge-projected contributions that are weighted differently due
to the SQD electron spin.

(iv) Finally, our results cover a broad experimentally relevant regime of
finite voltages and temperatures, and not only limits of, e.g., infinite bias
voltage $V_b$ in {\Cite{Gurvitz05,Gurvitz08}} or zero temperature $T$ as in
{\Cite{Shnirman98,Makhlin01a}}. The interplay of the above renormalization
effects leads to nontrivial voltage dependencies, similar to that in quantum
dot spin-valves.

On the technical side, we provide an important {\emph{sum rule}} for the
qubit dynamics: the kinetic equations must reproduce the free qubit evolution
(i.e., for zero tunnel coupling) when tracing over the \emph{interacting}
SQD degrees of freedom in addition to the electrodes. This is a concrete
application of the generalized current conservation law discussed in
{\Cite{Salmilehto12}}. We show that our kinetic equations are consistent with
this current conservation. It may be violated if instead a Born-Markov
approach followed by a secular approximation is applied{\cite{Salmilehto12}}
as we demonstrate for our concrete model in {\App{sec:secular}}. More
generally, such a sum rule has to hold for any observable that is conserved by
the tunneling. We furthermore prove in {\App{sec:sumRule}} that any kinetic
equation derived from real-time diagrammatics respects this sum rule
order-by-order in SQD tunnel coupling $\Gamma$.

Compared with previous works, however, our study is limited: we focus on the
analysis of the kinetic equations in the stationary limit. Although for
quantum information processing ultimately the measurement dynamics is of
interest, we \ apply our general kinetic equations only to test measurements
designed to verify that the SQD couples to a nearby qubit at all. We compare
the ensemble-averaged current and differential conductance through the SQD as
the readout strength is varied.{\todo{$\sim \Gamma \lambda$}} Our study
clearly indicates that already here the isospin torque terms have a
significant impact. This indicates that these terms will also influence the
transient behavior of the qubit in the measurement process. The kinetic
equations that we derive, however, provide a starting point for a more general
analysis of coherent backaction effects, which is, however, beyond the scope
of this paper. This is of interest both for understanding the limitations of
qubit readout devices as well as for exploring new means of controlling qubits
by coherent backaction effects.

The paper is organized as follows. After formulating the model in
{\Sec{sec:model}}, we introduce in {\Sec{sec:results}} the charge-projected
qubit isospins and analyze their dynamics in dependence on the SQD charge, in
particular the isospin-torque contributions to the kinetic equations. In
{\Sec{sec:signal}}, we illustrate the quantitative importance of the isospin
torques for the readout current through the sensor QD and study the
corrections it gives to the stationary read-out current and differential
conductance, which is often measured directly. We find that torque terms may
significantly alter the qubit-dependent \ conductance, up to 30\% for typical
experimental values for the asymmetry of the SQD tunnel couplings. We
summarize and discuss possible extensions in {\Sec{sec:summary}}.

{\Figure{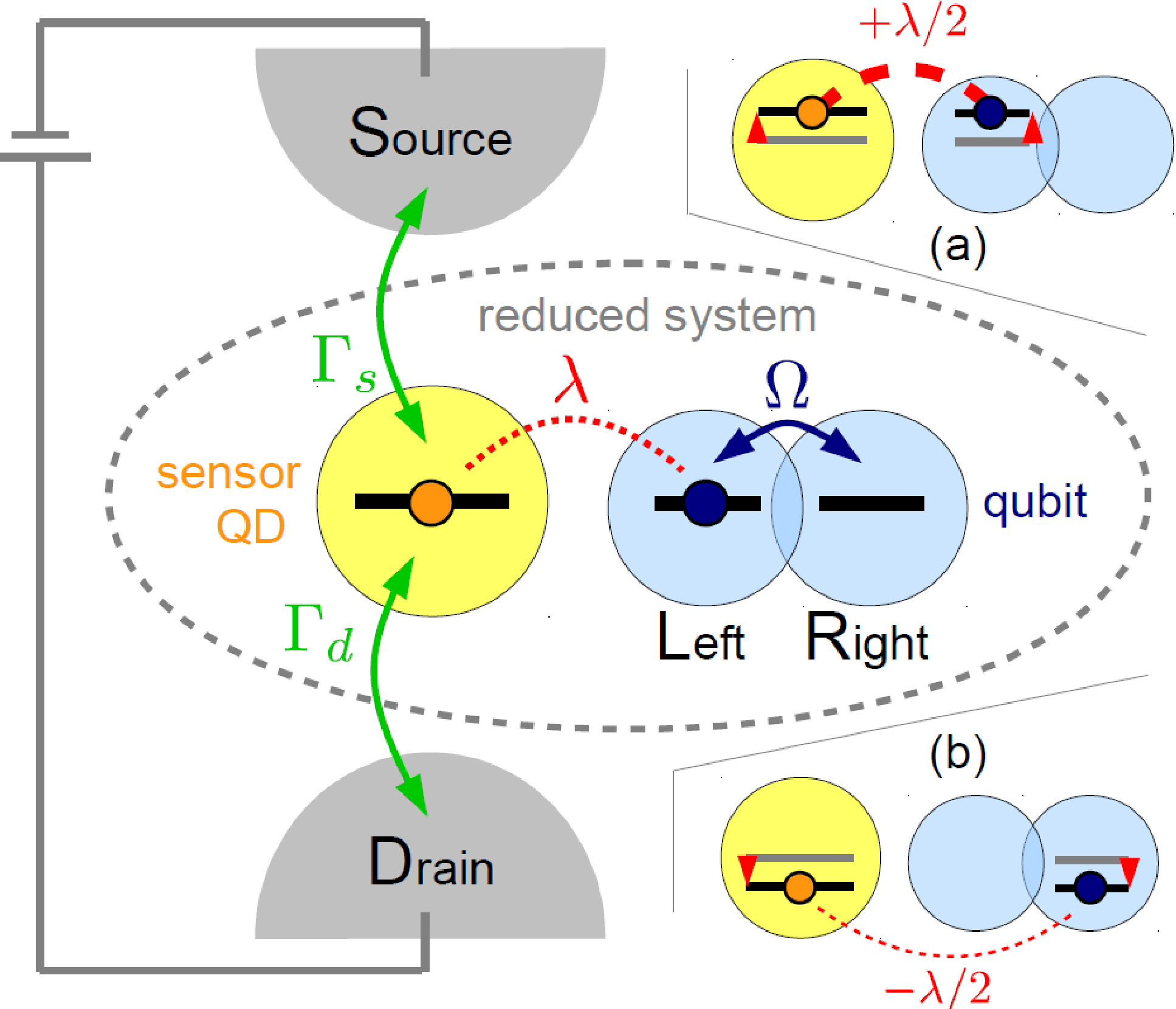}{Sensor quantum dot (SQD) tunnel-coupled to
source and drain electrode and capacitively coupled to a qubit, whose
different logical states involve two possible positions left and right in a
double quantum dot. If the qubit electron is left (a) or right (b), the
Coulomb repulsion to the SQD electron is larger or smaller, respectively,
compared to full delocalization of the qubit electron. We note that the
spin-qubit readout can be mapped onto a charge-qubit readout utilizing
spin-to-charge conversion.\label{fig:model} \ }}

\section{Qubit readout}\label{sec:model}

The readout of spin qubits is usually reduced to a charge readout by
spin-to-charge conversion {\cite{Elzerman04,Barthel10}}. Therefore, we focus
here on the conceptually clearest problem, sketched in {\Fig{fig:model}},
namely the capacitive readout of a charge-qubit. The qubit itself consists of
a double quantum dot occupied by a single electron, which occupies either an
orbital localized on the left or right dot ($l = L, R$). These states are
denoted by
\begin{eqnarray}
  \ket{l} & = & a_l^{\dag} \ket{0}_Q, 
\end{eqnarray}
with $a_l$, $a_l^{\dag}$ denoting the electron field operators of the qubit.
The qubit is described by a Hamiltonian accounting for a real hopping
amplitude $\Omega$ between the orbitals:
\begin{eqnarray}
  H_Q & = & \tfrac{1}{2} \Omega (a_L^{\dag} a_R + a_R^{\dag} a_L) . 
  \label{eq:HQ}
\end{eqnarray}
The isolated eigenstates of the qubit are thus superposition states
$\tfrac{1}{\sqrt{2}} (a_L^{\dag} \pm a_R^{\dag}) \ket{0}$, corresponding to an
isospin in the $x$ direction when $a_L^{\dag} \ket{0}$ and $a_R^{\dag}
\ket{0}$ denote isospin-up and -down along the $z$ direction, respectively.
Since we assume that the real spin of the qubit electron does not couple to
the measurement device, it remains fixed and we have omitted this quantum
number from the beginning.

The sensor quantum dot (SQD) is modeled by a single, interacting,
spin-degenerate orbital level with Hamiltonian
\begin{eqnarray}
  H_S & = & \sum_{\sigma} \varepsilon n_{\sigma} + U n_{\uparrow}
  n_{\downarrow}, 
\end{eqnarray}
containing the occupation operator $n_{\sigma} = d^{\dag}_{\sigma} d_{\sigma}$
for electrons of spin $\sigma = \uparrow, \downarrow$, whose annihilation and
creation operators are $d_{\sigma}$ and $d_{\sigma}^{\dag}$, respectively. Due
to the strong Coulomb repulsion of the quantum-confined electrons, the double
occupation of the SQD costs an additional energy $U$. Typically, $U$ is the
largest energy scale (except for the bandwidth of the leads, denoted by $D$):
charging energies are on the order of $\sim 0.1 - 1$ meV
{\cite{Fujisawa04,Lu03}}. Close to the SET resonance used for detection (cf.
{\Sec{sec:readout}}), this allows us to exclude the doubly occupied state of
the SQD, retaining only $\ket{0}_S$ and $\ket{\sigma}_S = d_{\sigma}^{\dag}
\ket{0}_S$. This, however, implies that we need to keep track of the spin
degree of freedom of the electrons (neglected in
{\Cite{Gurvitz08,Shnirman98,Makhlin01a}}) unless a high magnetic field is
applied. However, for singlet-triplet qubits the applied magnetic fields
(required to define the qubit) are in the range of $\sim 100 \mathrm{mT}$
{\cite{Barthel09,Petta05}}. The corresponding energy splittings in GaAs are a
few $\mu \mathrm{eV}$, which is much smaller than typical voltage bias $\sim 50
\mu \mathrm{eV}$ {\cite{Barthel10}}. Therefore, both spin channels are
energetically accessible and in general are relevant for the detection current
through the SQD with noticeable consequences, as we discuss in
{\Sec{sec:readout}}. Furthermore, these magnetic energies are of the same
order as typical tunneling rates $\Gamma$. This implies that renormalization
effects may be important since states of the SQD plus qubit system can be
mixed by the tunnel coupling to the electrodes. For the sake of simplicity, we
assume zero magnetic field here, resulting in energy-degenerate spin-up and
spin-down states on the SQD.

The readout of the qubit state using the SQD involves two couplings: the first
one is the capacitive interaction of the total charge $n = n_{\uparrow} +
n_{\downarrow}$ on the sensor QD with the charge configuration of the qubit,
given by
\begin{eqnarray}
  H_I & = & \tfrac{1}{2} n \lambda (a_L^{\dag} a_L - a_R^{\dag} a_R) . 
  \label{eq:Hint}
\end{eqnarray}
This qubit-dependent energy shift by $\pm \lambda / 2$ in turn affects the
charge current through the SQD from the source to the drain electrodes, which
are described as noninteracting reservoirs,
\begin{eqnarray}
  H_R & = & \sum_{r, k, \sigma} \omega_{r k \sigma} c^{\dag}_{r k \sigma} c_{r
  k \sigma},  \label{eq:reservoirs}
\end{eqnarray}
each in equilibrium at common temperature $T$, but held at different
electrochemical potentials $\mu_s = V \text{/} 2$ and $\mu_d = - V / 2$. Here,
$c_{r k \sigma}$ are the field operators referring to orbital $k$ and spin
$\sigma$ in source ($r = s$) and drain ($r = d$), respectively. The second
coupling involved in the readout process is the tunneling from the SQD to the
electrodes, and vice versa, accounted for by
\begin{eqnarray}
  H_T & = & \sum_{r, k, \sigma} t_r c^{\dag}_{r k \sigma} d_{\sigma} +
  \text{H.c.} .  \label{eq:HT}
\end{eqnarray}
The relevant energy scale is given, in terms of the tunneling amplitudes $t_r$
and the density of states $\nu_r$ of lead $r$, by the tunneling rates
$\Gamma_r = 2 \pi | t_r |^2 \nu_r$. For the sake of simplicity, we take both
$t_r$ and $\nu_r$ to be spin $(\sigma)$ and energy $(k)$-independent. The
source-drain coupling asymmetry $\gamma = \sqrt{\Gamma_s / \Gamma_d}$ of the
SQD, however, is a crucial parameter.

For our analysis, we will assume the conceptually simplest continuous
measurement limit $\lambda, \Omega \ll \Gamma$ and perform a controlled
perturbative calculation. In the experimental situation
{\cite{Lu03a,Barthel10}}, in which the coupling $\lambda \sim \Gamma$, the
effects may be even stronger.

\section{Charge-dependent isospin dynamics}\label{sec:results}

\subsection{Density operator and isospins}

In the following, we express the {\emph{action}} of the qubit state on the SQD
and the corresponding {\emph{backaction}} in terms of the isospin operator
\begin{eqnarray}
  \hat{\tau}_i & = & \sum_{l, l'} (\sigma_i)_{l l'} a^{\dag}_l a_{l'}, 
\end{eqnarray}
where $\sigma_i$ denotes the Pauli matrix for $i = x, y, z$. The ensemble
average of the isospin $\vecg{\tau} = \brkt{\op{\vecg{\tau}}}$ is obtained by
averaging over the state of (integrating out) \emph{both} electrodes
\emph{and} SQD. This qubit Bloch vector characterizes the reduced density
operator of the qubit and is conveniently normalized to 1. Its $z$ component
$\tau_z$ quantifies the imbalance of the probability to find the qubit
electron in left orbital rather than the right orbital, while $\tau_x$ and
$\tau_y$ quantify coherences between the left and right occupations.

It is, however, difficult to directly obtain a kinetic equation for the
isospin $\vecg{\tau}$ while incorporating the various effects of the
measurement. A general reason for this is that the SQD is also a microscopic
system, so that the action of the qubit on the sensor dynamics is not
negligible, which then in turn affects the backaction of the sensor on the
qubit. Another complication arises since we take into account an
\emph{interacting} detector (with proper spinful electrons), which
technically can not be integrated out easily. Moreover, the SQD is driven out
of equilibrium by the connected electron reservoirs.

We therefore instead derive a kinetic equation for the reduced density matrix
$\rho (t)$ of SQD {\emph{plus}} qubit by integrating out only the electrodes.
Yet, this requires \emph{two} Bloch vectors for a complete description of
the qubit, as we now explain. The Hilbert space of the joint qubit-SQD system
is spanned by six states denoted by $\ket{\sigma}_S \otimes \ket{l}_Q$ with $l
= L, R$ referring to the state of the qubit and $\sigma = 0, \uparrow,
\downarrow$ denoting the state of the SQD. Thus, the reduced density operator
$\rho (t)$ corresponds to a $6 \times 6$ matrix. However, since the charge,
the $z$ component of the (real) spin, and the total spin are conserved for the
total system including the leads {\cite{Leijnse08a}}, the reduced density
operator is diagonal in the SQD degrees of freedom and independent of the
choice of quantization axis of the real spin. Thus, we only need two $2 \times
2$ density matrices $\rho^n_Q$, one for each of the two charge states of the
SQD $n = 0, 1$:
\begin{eqnarray}
  \rho & = & \op{P}^0 \rho^0_Q + \op{P}^1 \rho^1_Q .  \label{eq:rhoExpand}
\end{eqnarray}
Here, $\op{P}^n$ denotes the operator projecting onto the charge state $n = 0,
1$ of the SQD, that is,
\begin{eqnarray}
  \op{P}^0 & = & \text{}_{} \ket{0}_S  \text{}_S \bra{0} \otimes
  \mathbbm{1}_Q,  \label{eq:P0}\\
  \op{P}^1 & = & \sum_{\sigma = \uparrow, \downarrow} \text{}_{}
  \ket{\sigma}_S  \text{}_S \bra{\sigma} \otimes \mathbbm{1}_Q,  \label{eq:P1}
\end{eqnarray}
where $\mathbbm{1}_Q = \sum_l \ket{l}_Q  \text{}_Q \bra{l}$ is the qubit unit
operator. Next, expanding each $\rho^n_Q$ in {\Eq{eq:rhoExpand}} in terms of
the unit and three Pauli matrices, we find that the relevant part of the
density operator is parametrized by only eight real expectation values $p^n =
\mathrm{tr} ( \hat{P}^n \rho)$ and $\tau^n_i = \mathrm{tr} ( \op{P}^n  \op{\tau}_i
\rho)$:
\begin{eqnarray}
  \rho & = & \tfrac{1}{2}  \sum_n p^n \op{P}^n + \tfrac{1}{2}  \sum_{n, i}
  \tau^n_i \left( \op{P}^n \op{\tau}_i \right) .  \label{eq:redDens}
\end{eqnarray}
The numbers $p^n = \mathrm{tr} ( \hat{P}^n \rho)$ give the probability for the
SQD to be in charge state $n = 0$ or $1$. Probability conservation is
expressed by
\begin{eqnarray}
  p^0 + p^1 & = & 1.  \label{eq:pcons}
\end{eqnarray}
Furthermore, $\tau^n_i$ are the averages of the isospin components $i = x$,
$y$, and $z$ when the SQD is in charge state $n = 0$ or $1$, respectively. By
definition these {\emph{charge-projected}} isospins sum up to the average of
the total isospin,
\begin{eqnarray}
  \vecg{\tau}^0 + \vecg{\tau}^1 & = & \vecg{\tau} . 
\end{eqnarray}
\subsection{Kinetic equations}\label{sec:kineq}

The Hamiltonian of the isolated reduced system (qubit plus SQD with $H_T = 0$)
can be expressed in terms of the isospin operator as
\begin{eqnarray}
  H_{\text{red}} & = & \tfrac{1}{2} \vec{\Omega} \cdot \op{\vecg{\tau}} +
  \op{P}^1  \left( \varepsilon + \tfrac{1}{2}  \vecg{\lambda} \cdot
  \op{\vecg{\tau}} \right)  \label{eq:Hred}
\end{eqnarray}
where the effective magnetic fields of the qubit mixing, $\vec{\Omega} =
\Omega \vec{e}_x$, and the readout, $\vecg{\lambda} = \lambda \vec{e}_z$, are
orthogonal. Here we see the \emph{action} of the measurement: the state of
the qubit modulates the effective level position of the SQD between
$\varepsilon \pm \lambda \text{/} 2$. This affects the energy-dependent
tunneling rates between the SQD and the leads and by this the measurable
transport current.

{\color{blue} }

The kinetic equations for the isolated reduced system, obtained from the
von Neumann equation $\dot{\rho} = - i [H_{\mathrm{red}}, \rho]$, show that the
charge-projected isospins are subject to different, noncollinear effective
magnetic fields $\vec{\Omega}$ and $\vec{\Omega} + \vecg{\lambda}$:
\begin{eqnarray}
  \dot{p}^0 & = & \dot{p}^1 = 0,  \label{eq:p0free}\\
  \dot{\vecg{\tau}}^0 & = & \vec{\Omega} \times \vecg{\tau}^0, 
  \label{eq:tau0free}\\
  \dot{\vecg{\tau}}^1 & = & \left( \vec{\Omega} + \vecg{\lambda} \right)
  \times \vecg{\tau}^1 .  \label{eq:tau1free}
\end{eqnarray}
If the SQD is singly occupied, the capacitive interaction $\vecg{\lambda}$
thus exerts a \emph{backaction} torque, perturbing the free qubit isospin
precession about $\vec{\Omega}$.

We note that the ensuing analysis is, in general, not limited to charge
qubits: For example, in singlet-triplet qubits, two exchange-coupled electrons
in the qubit double quantum dot form spin-singlet and spin-triplet states,
which due to a different charge distribution also couple capacitively to a
sensor quantum dot (spin-to-charge conversion). However, the two-electron
double quantum dot Hilbert space is {\emph{four}}-dimensional instead of
two-dimensional as for the charge qubit, which introduces an additional
complexity to the problem that is beyond the scope of this paper. Still, as
long as two electrons stay in the qubit subspace formed by the spin-singlet
and the spin-triplet state $T_0$, the Hamiltonian {\eq{eq:Hred}} provides a
valid model for the readout of a singlet-triplet qubit if one included a $z$
component into $\mathbf{\vec{\Omega}}$, accounting for the exchange
splitting $J$ between $S$ and $T_0$ (the other two triplet states $T_+$ and
$T_-$ are usually energetically split off by a large real magnetic field $B
\gg J$). Note that the exchange interaction $J$ between the electrons is
typically in the order of a few $\mu$eV {\cite{Petta05}}, which can be well
below the tunneling rate $\Gamma$. This is a crucial requirement for our
analysis.

When including the tunnel coupling $H_T$ of the SQD to the electronic
reservoirs, Eqs. {\eq{eq:p0free}}--{\eq{eq:tau1free}} turn into a set of
equations that couple the occupation probabilities $p^n$ of the SQD and the
charge-projected isospins $\tau^n$ ($n = 0, 1$). We derive these from the
kinetic equation for the reduced density operator using the real-time
diagrammatic technique {\cite{Schoeller94,Schoeller09a,Leijnse08a}}, including
all coefficients of order $\Gamma$ as well as $\Gamma^2 \text{/} T$,
$\lambda \Gamma \text{/} T $, and $\Omega \Gamma \text{/} T$ and
neglecting remaining terms of higher order in $\Gamma$, $\lambda$, and
$\Omega$. In addition we make a Markov approximation. In {\App{sec:RTD}}, we
explain how to perform this expansion and justify its validity under the
conditions $\lambda, \Omega \ll \Gamma \ll T$. 
Within these approximations the
kinetic equations read as

{\wideeq{\begin{eqnarray}
  \tfrac{d}{d t} \left(\begin{array}{c}
    p^0\\
    p^1\\
    \vecg{\tau}^0\\
    \vecg{\tau}^1
  \end{array}\right) & = & \left(\begin{array}{cccc}
    - 2 \Gamma^0 & + \Gamma^1 & + 2 \vec{C} \cdot & + \vec{C} \cdot\\
    + 2 \Gamma^0 & - \Gamma^1 & - 2 \vec{C} \cdot & - \vec{C} \cdot\\
    + 2 \vec{C} & + \vec{C} & - 2 \Gamma^0 + \left( \vec{\Omega} - 2
    \vecg{\beta} \right) \times & \Gamma^1 - \vecg{\beta} \text{} \times\\
    - 2 \vec{C} & - \vec{C} & + 2 \Gamma^0 + 2 \vecg{\beta} \times & -
    \Gamma^1 + \left( \vec{\Omega} + \vecg{\lambda} + \vecg{\beta} \right)
    \times
  \end{array}\right) \left(\begin{array}{c}
    p^0\\
    p^1\\
    \vecg{\tau}^0\\
    \vecg{\tau}^1
  \end{array}\right) .  \label{eq:kinEq}
\end{eqnarray}}}
When computing the matrix product with the column vector in the above equation, the dot ``$\cdot$'' (cross ``$\times$'') in the entries of the matrix indicates that a three-dimensional scalar (vector) product is to be formed.
The coefficients in \ {\Eq{eq:kinEq}} are $\Gamma^n = \sum_{r = s, d}
\Gamma^n_r$ with the \emph{renormalized} dissipative SQD rates through
junction $r = L, R$,
\begin{eqnarray}
  \Gamma^{0, 1}_r & = & \Gamma_r f_r^{\pm} \pm \sum_{q = s, d} \tfrac{\Gamma_r
  \Gamma_q}{2 T} (f^+_r)' \phi_q \nonumber\\
  &  & \mp \sum_{q = s, d} \tfrac{\Gamma_r \Gamma_q}{2 T} \phi'_r (2 f^+_q +
  f^-_q),  \label{eq:dissRate}
\end{eqnarray}
the vector $\vec{C} = \sum_r \vec{C}_r$ with the isospin-to-charge conversion
rates
\begin{eqnarray}
  \vec{C}_r & = & \tfrac{\Gamma^r}{2 T}  \vecg{\lambda}   (- f_r^+
  )',  \label{eq:conversion}
\end{eqnarray}
which are vectors with positive elements, and finally the vector $\vecg{\beta}
= \sum_r \vecg{\beta}_r$, giving rise to isospin torque terms, with the
effective magnetic fields
\begin{eqnarray}
  \vecg{\beta}_r & = & \tfrac{\Gamma_r}{T} \left( \vec{\Omega} + \tfrac{1}{2}
  \vecg{\lambda} \right) \phi_r' .  \label{eq:exField}
\end{eqnarray}
In {\Fig{fig:rates}}, we plot the contribution of a single electrode to the
magnitude of these coefficients as a function of the gate voltage $V_g$. In
the above expressions $f^{\pm}_r = f^{\pm} ((\varepsilon - \mu^r) \text{/} T)$
abbreviates the Fermi function for lead $r = s, d$ with $f^+ (x) = (e^x -
1)^{- 1}$ and $f^- (x) = 1 - f^+ (x)$. The level renormalization function
$\phi_r = \phi ((\varepsilon - \mu^r) \text{/} T)$ is defined by
\begin{eqnarray}
  \phi (x) & = & \mathcal{P}  \int_{- \Lambda}^{+ \Lambda} \frac{d y}{\pi} 
  \frac{f^+ (y)}{x - y}  \label{eq:phi}\\
  & = & - \mathrm{Re} \text{ } \psi \left( \frac{1}{2} + i \frac{x}{2 \pi}
  \right) + \ln \left( \frac{\Lambda}{2 \pi} \right) . 
\end{eqnarray}
Derivatives are indicated by a dash: $\phi' = \partial \phi (x) \text{/}
\partial x$ and analogously $(f^+)' = \partial f^+ (x) \text{/} \partial x$.
In {\Eq{eq:phi}}, $\mathcal{P} $ denotes the principal value of the integral
with a cutoff $\Lambda = D \text{/} T$, yielding the real part of the digamma
function $\psi$ and a logarithmic correction. The latter depends on the
electrode bandwidth $D$, which must be set to $D \sim U$ where $U$ is the
large local Coulomb interaction energy of the SQD (we excluded the doubly
occupied state from the SQD Hilbert space). In this way $D \sim U$ enters into
the rates, {\Eq{eq:dissRate}}, but $D$ drops out in the derivatives required
for the torque terms, {\Eq{eq:exField}}. Finally, although we refer to
{\Eq{eq:dissRate}} simply as the ``renormalized'' SQD rate, one should note
that the $O (\Gamma^2)$ corrections to the $O (\Gamma)$ rate (first term)
includes both renormalization of the energy level $\varepsilon$ (second term)
as well as an elastic cotunneling contribution (third term).
{\Fig{fig:rates}}(a) shows that this leads to a combined shift and broadening
of the resonant step in the rates around $\varepsilon \approx \mu_r$; cf.
{\App{sec:cot}}. We note that we were careful to restrict
our study to weak couplings $\Gamma \ll T$. This clearly validates the neglect of even
higher-order corrections. In particular, we can safely exclude the
occurrence of Kondo
physics even for those results we show in the Coulomb
blockade regime.

{\Figure{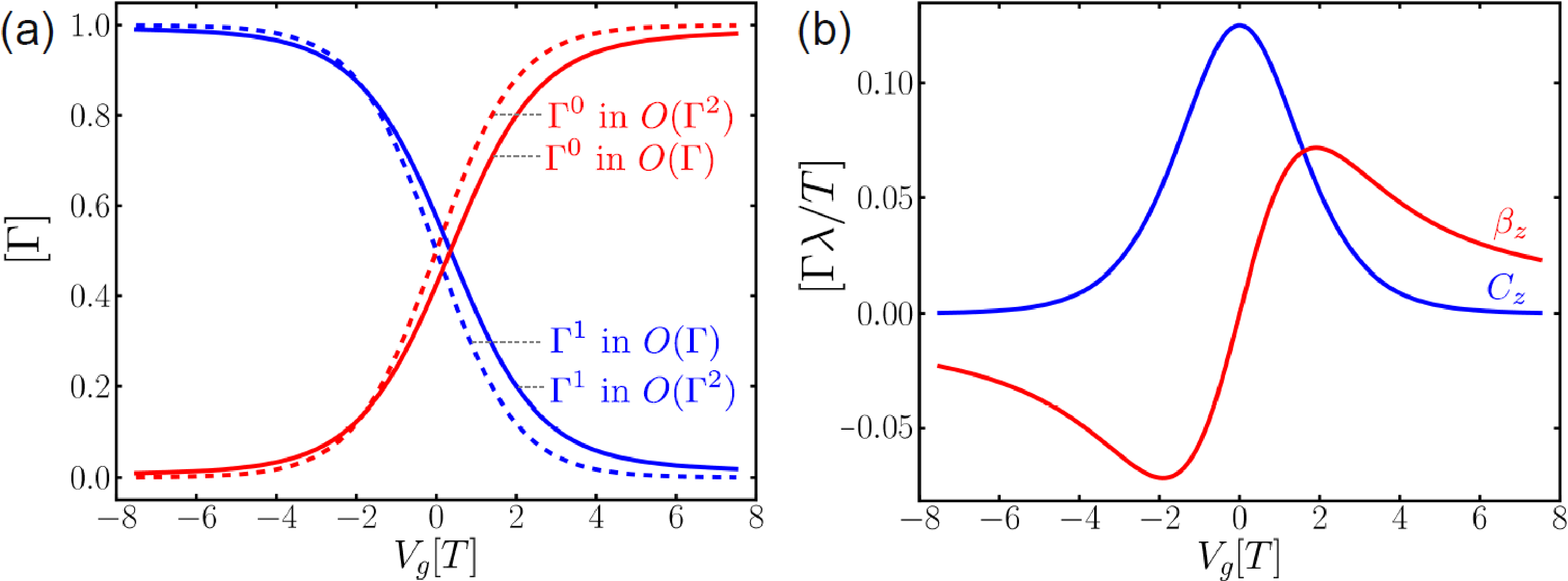}{Dissipative and coherent coefficients induced by
a single electrode for $\mu = 0$ (electrode index suppressed for simplicity).
(a) Dissipative SQD rates $\Gamma^{0, 1}$ {\eq{eq:dissRate}} including the
renormalization of the tunneling (solid lines) and excluding it (dashed lines)
as a function of gate voltage $V_g = - \varepsilon$. Parameters: $\Gamma
\text{/} T = 0.2$ and $D / T = 10^2$. (b) $z$ component of the
isospin-to-charge conversion rate $C_z$, \ {\Eq{eq:conversion}}, (blue) and
the isospin torque $\beta_z$, {\Eq{eq:exField}}, (red) as a function of $V_g$.
Since $\Omega_z = 0$ the curve with the scaling chosen is independent of
further parameters. Note that the $C_z$ drops exponentially in the Coulomb
blockade regime, whereas $\beta_z$ decreases only
algebraically.\label{fig:rates}}}

\subsection{Sum rules}

The kinetic equations {\eq{eq:kinEq}} clearly satisfy the sum rule $\dot{p}^0
+ \dot{p}^1 = 0$, which expresses the probability conservation,
{\Eq{eq:pcons}}. Moreover, we discuss in {\App{sec:sumRule}} that a second sum
rule exists for the charge-projected isospins: their sum has to reproduce the
{\emph{internal}} evolution of the \emph{total average} isospin, i.e. as
if the tunneling was switched off ($H_T = 0$); see also
{\Cite{Salmilehto12}}. Their sum is thus given by Eqs.
{\eq{eq:tau0free}} and {\eq{eq:tau1free}}:
\begin{eqnarray}
  \dot{\vecg{\tau}}^0 + \dot{\vecg{\tau}}^1 & = & \dot{\vecg{\tau}}
  |_{\text{int}} =   \vec{\Omega} \times \left( \vecg{\tau}^0 +
  \vecg{\tau}^1 \right) + \vecg{\lambda} \times \vecg{\tau}^1 . 
  \label{eq:tausum}
\end{eqnarray}
This constrains the dynamics of the average charge-projected isospins
$\vecg{\tau}^0$ and $\vecg{\tau}^1$, without reducing one to the other (as
happens for $p^0$ and $p^1$). The isospin sum rule is a consequence of the
conservation of the total isospin \emph{operator} in the tunneling, that
is, $\left[ \op{\vecg{\tau}}, H_T \right] = 0$ {\cite{Salmilehto12}}. It holds
in the presence of the reservoirs, order-by-order in the perturbation
expansion in $\Gamma$. Indeed, we find that our kinetic equations obey
{\Eq{eq:tausum}}, as do the results in {\Cite{Makhlin01a}}. By contrast, Eqs. (31a)-(31f) given in {\Cite{Gurvitz08}} in general violate it, unless
one expands to lowest order in the SQD-qubit coupling $\lambda$. In that case,
assuming energy-dependent tunneling rates, they agree with our kinetic
equations if we (i) send the bias to infinity, implying energy-independent
Fermi functions $f_L^+ = f_R^- = 1$ and $f_L^- = f_R^+ = 0$ in
{\Eq{eq:kinEq}}, (ii) neglect all renormalization effects, i.e., the
isospin torque and the renormalization of the tunneling rates, and (iii)
ignore factors of two due to the SQD spin (importance discussed in
{\Sec{sec:readout}}).

\subsection{Isospin torques}

The kinetic equations {\eq{eq:kinEq}} together with the probability
conservation $p^0 + p^1 = 1$ completely determine the Markovian dynamics of
the reduced SQD-qubit system in the limit{\footnote{If we set $\lambda = 0$ in
{\Eq{eq:kinEq}}, the resulting equations for the occupations and the
charge-projected isospins decouple. The equations for the occupations coincide
with those for the $U = \infty$ Anderson model up to order $\Gamma^2$ (i.e.,
the SQD without the qubit present). Furthermore, when integrating out the SQD,
we reproduce the dynamics of a freely evolving qubit: $\dot{\vecg{\tau}} =
\dot{\vecg{\tau}}^0 + \vecg{\tau}^1 = \vec{\Omega} \times \vecg{\tau}$.
Notably, this equation does not depend on isospin torque terms which, for
nonzero $\Omega$, still remain in {\Eq{eq:kinEq}} for $\lambda = 0$. Despite
their appearance, these terms thus have no physical consequence in this limit,
as it should be. We furthermore note that there is no unique stationary state
of the {\emph{joint}} SQD-qubit system in both the cases $\lambda = 0$ and
$\Omega = 0$. This is expected since in these cases we completely decouple a
subsystem. Finally, we note that if we formally set $\Gamma = 0$, we recover
the free evolution of Eqs. {\eq{eq:tau0free}} and {\eq{eq:tau1free}}.}}
$\lambda, \Omega \ll \Gamma \ll T$. The dynamics of the occupations and the
isospins are coupled by the charge-to-isospin conversion vector rates
$\vec{C}$: in the first line of {\Eq{eq:kinEq}}, they describe the influence of
the qubit state on the occupations, that is, the measurement action, which
modifies the current, whereas in the second and third lines of {\Eq{eq:kinEq}},
they represent a \emph{backaction} of the measurement on the qubit. These
\emph{dissipative} terms scale as $\sim \Gamma \lambda \text{/} T$, i.e.,
with the product of {\emph{both}} couplings that are involved in the
measurement process.

When keeping the above terms $\vec{C} \sim \Gamma \lambda \text{/} T$ that
describe the readout action and backaction, we \emph{must} also keep
torque terms $\vecg{\beta} \times \vecg{\tau}^n$ induced by the readout since
$\vecg{\beta}$ scales in the same way (since $\lambda \sim \Omega$ or even
$\lambda < \Omega$) unless prefactors are very small. These torque terms
represent a \emph{coherent} backaction on the qubit since it derives from
level renormalization effects. The isospin torque terms have a nontrivial
voltage dependence. At the resonance of the SQD level with the electrochemical
potential $\mu_r$, $\varepsilon = - V_g = \mu_r$, the effective magnetic field
from lead $r$ vanishes, $| \vecg{\beta}_r | = 0$. However, it sharply rises to
two extrema at $| \varepsilon - \mu_r | \approx 2 T$, i.e., at the ``flanks''
of the Coulomb peaks. Figure {\fig{fig:rates}}(b) shows that here $| \vecg{\beta} |
\sim | \vec{C} |$, right at the crossover regime from single electron
tunneling to Coulomb blockade where the SQD has the highest readout
sensitivity. Further away from resonance the dissipative (back)action terms
($\vec{C}$) are exponentially suppressed with $| \varepsilon - \mu_r |$, so
that the torque terms even start to dominate: the field $\vecg{\beta}$ only
decays algebraically with $\phi'_r \sim 1 \text{/$| \varepsilon - \mu_r |$}$.
The latter approximation holds when the bias is the largest energy scale.

It is explicitly stated in {\Cite{Makhlin01a,Shnirman98}} that level
renormalization contributions are neglected. 
In the limit of infinite bias
$V_b$ torque effects can be neglected as done in
{\Cites{Gurvitz05,Gurvitz08}} because the magnitude of $\vecg{\beta}$ scales as
$\left| \vecg{\beta} \right| \sim 1 \text{/} V_b$.
Thus, our results agree in this regard with
{\Cites{Gurvitz05,Gurvitz08}}. Altogether, it is therefore not
surprising that the coherent backaction has not been noted so far.
However, if the bias $V_b$ is large, but finite,
corrections from renormalization effect can still be sizable, see also
{\Cite{Wunsch05}} for a related discussion. Thus, one
should also reckon with renormalization effects when suppressing the readout
current by tuning the SQD into Coulomb blockade where the qubit state is
supposed {\emph{not}} to be measured (during other processing steps, e.g.,
qubit manipulation). Although this nearly eliminates the dissipative
backaction ($\left| \vec{C} \right| \approx 0$), the coherent backaction may
still be of appreciable size. Thus, this nontrivial dependence of the induced
torque on the gate (and bias) voltage, illustrated in {\Fig{fig:rates}}(b),
presents interesting possibilities that may be used to control the quantum
state of a qubit.

We finally discuss the relaxation rates $\sim \Gamma + \Gamma^2 \text{/} T$
[{\Eq{eq:dissRate}}]: they only contain the tunneling, i.e., they are
associated with the stochastic switching of the detector. Clearly, when
describing the readout for $\lambda \lesssim \Gamma$ one has to consistently
include the renormalization of the tunneling rates $\sim \Gamma^2 \text{/} T$
[second and third term in {\Eq{eq:dissRate}}]. To our knowledge, this has not
been pointed out so far (cf.
{\Cite{Gurvitz05,Gurvitz08,Shnirman98,Makhlin01a}}). In
{\App{sec:positivity}}, we show that this consistency is mandatory to obtain
physically meaningful results: when failing to account for the next-to leading
order terms $\sim \Gamma^2 \text{/} T$, while keeping the {\emph{coherent}}
backaction effects (isospin torques), the solution of the above kinetic
equations may severely violate the positivity of the density matrix
{\eq{eq:redDens}}.

\subsection{Analogy to quantum-dot spin-valves}\label{sec:analogy}

{\Figure{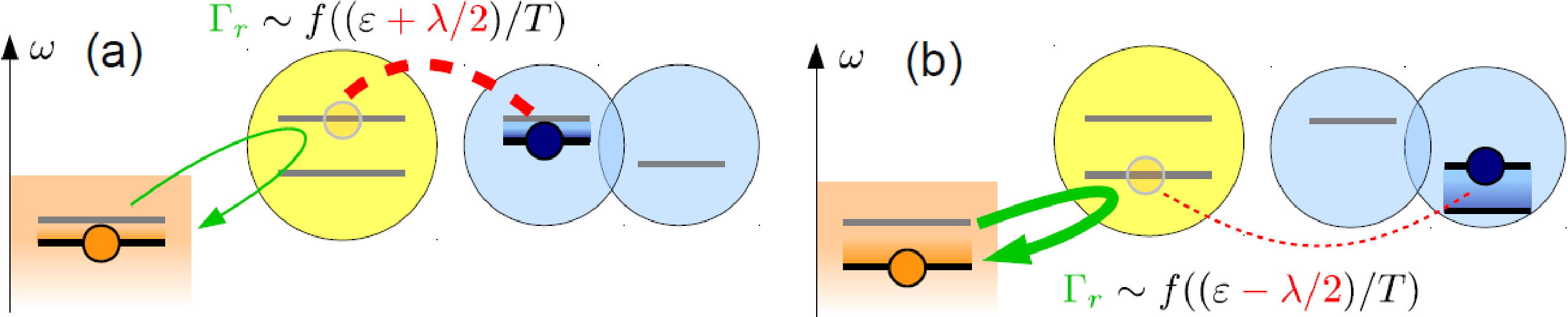}{Illustration of coherent processes
responsible for the torque terms $\sim \vecg{\beta}$ in {\Eq{eq:kinEq}} for
$\Omega = 0$ (for further explanation, see {\Sec{sec:analogy}}).
\label{fig:renormalization}}}

The torque terms in {\Eq{eq:kinEq}} are generated by coherent fluctuations of
electrons tunneling between the SQD and the leads. A similar coherent effect
is well-known in spintronics:{\cite{Braun04set,Koenig03}} when attaching a
quantum dot to ferromagnetic leads, an imbalance of the tunneling rates for
spin-up and spin-down electrons leads to a different level shift for the spin-up state and the spin-down state on the quantum dot. The resulting level splitting shows up
as an additional spin torque proportional to $\phi (\varepsilon)$ [see
{\Eq{eq:phi}}] in the kinetic equations resembling our qubit isospin
equations.{\cite{Braun04set}} In analogy to this case, here the strength of
the virtual fluctuations of the SQD depends on the position of the electron in
the qubit (see {\Fig{fig:renormalization}}). For the special case $\Omega = 0$
a simple argument can be made: if the isospin is up (down), the effective
level position of the qubit is shifted by $+ \lambda \text{/} 2$ ($- \lambda
\text{/} 2$). This gives a level splitting $\sim \phi_r \left( \tfrac{1}{T}
(\varepsilon + \tfrac{\lambda}{2}) \right) - \phi_r \left( \tfrac{1}{T}
(\varepsilon - \tfrac{\lambda}{2}) \right) \approx \tfrac{\lambda}{T} \phi'_r
(\varepsilon)$ in the weak coupling limit and explains why {\Eq{eq:exField}}
contains a {\emph{derivative}} of the renormalization function. [The factor
1/2 in {\Eq{eq:exField}} occurs because the isospin operator is not
normalized.] For nonzero mixing $\Omega$ of the qubit orbitals the additional
vector $\sim \tfrac{\vec{\Omega}}{T} \phi'_r (\varepsilon)$ appears in
{\Eq{eq:exField}} along a different direction. This accounts for a
renormalization of the qubit splitting.

These torque terms act on the charge-projected isospins,{\footnote{These terms
in {\Eq{eq:kinEq}} should not be confused with the mixing terms in Eqs.
(20)--(27) in {\Cite{Shnirman98}}. The latter equations contain matrix elements
of the reduced density operator with respect to the {\emph{different}}
eigenbases of the qubit depending on the charge state of the SQD. In contrast,
$\vecg{\tau}^0$ and $\vecg{\tau}^1$ in {\Eq{eq:kinEq}} refer both to the
{\emph{same}} (arbitrary) quantization axis. The mixing terms of
{\Cite{Shnirman98}} are contained in our charge-to-isospin conversion terms,
which is becomes clear when comparing to Eq. (5.10) in {\Cite{Makhlin01a}},
which rewrites the result in {\Cite{Shnirman98}}. }} but for the different
charge sectors they have opposite directions [as required by the sum rule
{\eq{eq:tausum}}] and differ in strength by a factor of 2 (due to the SQD
electron spin):
\begin{eqnarray}
  \dot{\vecg{\tau}}^0 & = & \left( \vec{\Omega} - 2 \vecg{\beta} \right)
  \times \vecg{\tau}^0 - \vecg{\beta} \text{} \times \vecg{\tau}^1 + \ldots 
  \label{eq:tau0dot},\\
  \dot{\vecg{\tau}}^1 & = & 2 \vecg{\beta} \times \vecg{\tau}^0 + \left(
  \vec{\Omega} + \vecg{\lambda} + \vecg{\beta} \right) \times \vecg{\tau}^1 +
  \ldots .  \label{eq:tau1dot}
\end{eqnarray}
As in kinetic equations for QD spin valves with nonzero spin in two adjacent
charge states {\cite{Baumgaertel11}}, in {\Eq{eq:tau0dot}}-{\eq{eq:tau1dot}}
we have a spin torque that precesses the isospin ($\dot{\vecg{\tau}}^0 \sim -
2 \vecg{\beta} \times \vecg{\tau}^0$ and $\dot{\vecg{\tau}}^1 \sim
\vecg{\beta} \times \vecg{\tau}^1$). However, in contrast to the spin-valve
equations, their sign is opposite for $n = 0$ and $n = 1$. Additionally, there
are torque terms that \emph{couple} the two isospins of different charge
states of the SQD ($\dot{\vecg{\tau}}^0 \sim - \vecg{\beta} \times
\vecg{\tau}^1$ and $\dot{\vecg{\tau}}^1 \sim 2 \vecg{\beta} \times
\vecg{\tau}^0$). Those turn out to be crucial as we explain in the next
section.

\section{Impact of coherent backaction}\label{sec:signal}

\subsection{Readout current}\label{sec:readout}

By taking the SQD spin and the strong local interaction $U$ on the SQD into
account, the dissipative rates involving $p^0$ and $\vecg{\tau}^0$ in our
kinetic equations {\eq{eq:kinEq}} differ from
{\Cite{Gurvitz05,Gurvitz08,Shnirman98,Makhlin01a}} by a factor of 2, as
expected.{\footnote{If the dot is empty, electrons from both spin channels can
enter the dot. This doubles the tunneling rates compared to the case when the
quantum dot is already occupied: then the residing electron can only leave the
dot into a single spin channel.}} A less obvious, but important difference
arises for the qubit-dependent part of the current flowing through the SQD,
i.e., the difference of the current for finite coupling $(I_{\lambda})$ and
zero coupling $(I_0)$
\begin{eqnarray}
  \Delta I & \assign & I_{\lambda} - I_0 . 
\end{eqnarray}
In the stationary limit ($\dot{p}^n = 0 ^{}$, $\dot{\vecg{\tau}}^n =
0$) $\Delta I$ may be expressed as
\begin{eqnarray}
  \Delta I & = & (\Gamma^0_s - \Gamma^0_d) (p^0_{\lambda} - p^0_0) -
  \tfrac{1}{2} (\Gamma^1_s - \Gamma^1_d) (p^1_{\lambda} - p^1_0) \nonumber\\
  &  & - \tfrac{1}{2} \left( \vec{C}_s - \vec{C}_d \right) \cdot \left( 2
  \vecg{\tau}^0_{\lambda} + \vecg{\tau}^1_{\lambda} \right) . 
  \label{eq:deltaI}
\end{eqnarray}
where $p^n_{\lambda}$, $\vecg{\tau}^n_{\lambda}$ and $p^n_0$,
$\vecg{\tau}^n_0$ are the stationary solutions of {\Eq{eq:kinEq}} for finite
and zero coupling $\lambda$, respectively. Although we will refer to $\Delta
I$ as the ``signal current'', {\Eq{eq:deltaI}} reveals it will in general
{\emph{not}} directly measure the position of qubit electron for two reasons.
First, the SQD occupations $p^n_{\lambda}$ depend on the isospins through the
kinetic equations [see {\Eq{eq:kinEq}}]: the last term term in
{\Eq{eq:deltaI}}, the one explicitly depending on the
$\vecg{\tau}^n_{\lambda}$, is not the only contribution to $\Delta I$.
Secondly, even this latter term is {\emph{not}} directly sensitive to the
total isospin $\vecg{\tau} = \vecg{\tau}^0 + \vecg{\tau}^1$ since
$\vecg{\tau}^0$ is weighted with the factor of 2 (due to the SQD-spin
degeneracy) relative to $\vecg{\tau}^1$ in {\Eq{eq:deltaI}}.

We now study the impact of the torque terms $\vecg{\beta} \times
\vecg{\tau}^n$ in {\Eq{eq:kinEq}} on the signal current $\Delta I$. Solving
{\eq{eq:kinEq}} in the stationary limit and inserting it into {\Eq{eq:deltaI}}
yields, to leading orders of $\lambda$ and $\Omega$,
\begin{eqnarray}
  \Delta I_{\beta = 0} & = & \frac{\left| \vec{C} \right|^2}{\Gamma^1} 
  \frac{\Gamma^0 + \Gamma^1}{2 \Gamma^0 + \Gamma^1} \left[ \frac{\left|
  \vec{C}_s \right|}{\left| \vec{C} \right|} - \frac{2 \Gamma^0_s +
  \Gamma^1_s}{2 \Gamma^0 + \Gamma^1} \right],  \label{eq:signal}\\
  \frac{\Delta I_{\beta \neq 0}}{\Delta I_{\beta = 0}} & = & 1 + \frac{\kappa
  (\Gamma^0 \text{/} \Gamma^1 - \Gamma^1 \text{/} \Gamma^0)}{1 - \kappa
  \text{/} 2 (1 - \Gamma^1 \text{/} \Gamma^0 + 2 \Gamma^0 \text{/} \Gamma^1)},
  \label{eq:signalTorque}
\end{eqnarray}
with $\kappa = \sum_r \left( \Gamma_r \text{$\phi'_r$/} T \right)$. At least
to lowest order, Eqs. {\eq{eq:signal}} and {\eq{eq:signalTorque}} are
independent of $\Omega$ [explained below {\Eq{eq:ansatz}}]. The signal
currents {\eq{eq:signal}} and {\eq{eq:signalTorque}} are plotted in
{\Fig{fig:deltaI}}(a); let us first focus on their main features. To this end
we neglect the change in the occupations of the SQD due to the coupling to the
qubit: setting $p^n_{\lambda} = p_0^n$ in {\Eq{eq:deltaI}} we obtain, for
symmetric tunnel couplings,
\begin{eqnarray}
  \Delta I & \approx & \tfrac{\Gamma \lambda}{4 T}  [(f_s^+)' - (f_d^+)'] (2
  \tau^0_z + \tau^1_z) .  \label{eq:deltaIEstimate}
\end{eqnarray}
In this case, a nonzero isospin polarization acts as an additional gate voltage
on the SQD and shifts the effective level position in the SQD to $\varepsilon
+ \lambda \tau^z \text{/} 2$ [see {\Eq{eq:Hred}}]. The signal current
{\eq{eq:deltaI}} is then just the linear response of the tunneling rates
$\Gamma_r^n (\varepsilon) \rightarrow \Gamma_r^n (\varepsilon + \lambda \tau^z
\text{/} 2)$ to that shift. In our case, the energy dependence of these rates
{\eq{eq:dissRate}} is mostly through the Fermi functions, which change sharply
when the level is aligned with the electrochemical potentials of source and
drain. This explains the s-shaped curve with a maximum / minimum roughly
expected at $V_g \approx \mp V_b \text{/} 2$, which is in fact slightly
shifted towards the adjacent Coulomb blockade regimes (see {\Fig{fig:deltaI}})
since $\tau^z$ still increases on the threshold to Coulomb blockade.

{\Figure{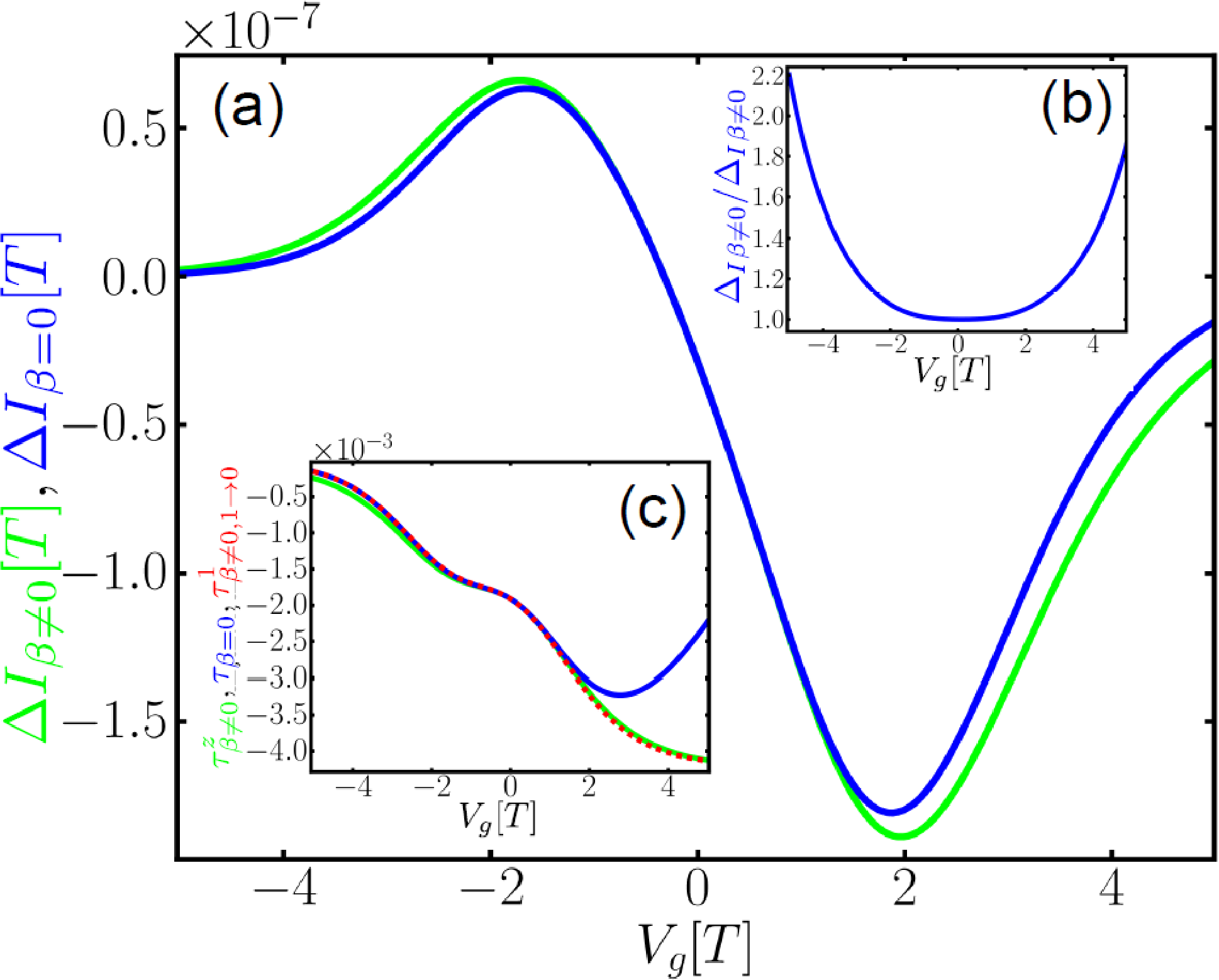}{(a, main panel) Signal current including torque
terms (green) and excluding torque terms (blue), and (b) the ratio of these two
currents as a function of gate voltage $V_g$. (c) $z$ component of the
{\emph{total}} isospin $\vecg{\tau} = \vecg{\tau}^0 + \vecg{\tau}^1$ including
torque terms (green), excluding torque terms (blue) and the $z$ component of
charge-projected isospin $\vecg{\tau}^1$ when we only keep the term
$\dot{\vecg{\tau}}^0 = - \vecg{\beta} \times \vecg{\tau}^1$ in {\Eq{eq:kinEq}}
(dashed red). In (a)-(c) the remaining parameters are the same: $V_b \text{/}
T = 3$, $\Gamma^L \text{/} T = \Gamma^R \text{/} T = 10^{- 1}$, $\lambda
\text{/} T = 10^{- 2}$, and $D / T = 10^3$.\label{fig:deltaI}}}

The second feature of {\Fig{fig:deltaI}}(a) is the notable asymmetry of the
s-shaped curve: the amplitude of the signal is larger for positive than for negative
gate voltages, which we explain in the following. To this end, we now first
neglect the torque terms in {\Eq{eq:kinEq}}. The stationary solution of the
resulting equations shows that the charge-projected isospins relax until they
are antiparallel to the effective field in that charge sector (the reduced
system tends to occupy the ground state), that is,
\begin{equation}
  \begin{array}{lllllll}
    \vecg{\tau}^0_{\beta = 0} & = & - c^0 \vec{\Omega}, &  &
    \vecg{\tau}^1_{\beta = 0} & = & - c^1 \left( \vec{\Omega} + \vecg{\lambda}
    \right),
  \end{array} \label{eq:ansatz}
\end{equation}
with $c^0, c^1 > 0$. Clearly, only $\vecg{\tau}^1_{\beta = 0}$ has a component
along the detection vector $\vecg{\lambda}$ and it therefore solely determines
\ the signal in the case of $\vecg{\beta} = 0$ by {\Eq{eq:deltaIEstimate}}.
When the gate voltage is lowered, the SQD is more likely to be empty and
$\vecg{\tau}^1_{\beta = 0}$ is suppressed. This is evident from the kinetic
equations {\eq{eq:kinEq}} since the relaxation rate $\Gamma^1$ of
$\vecg{\tau}^1_{\beta = 0}$ rises while the relaxation rate $\Gamma^0$ of
$\vecg{\tau}^0_{\beta = 0}$ becomes smaller when $V_g$ is lowered, thus
transferring a nonzero total isospin to the projection $\vecg{\tau}^0_{\beta =
0}$ rather than to $\vecg{\tau}^1_{\beta = 0}$. In conclusion, the signal has
the overall tendency to be decreased with $V_g$, explaining the asymmetry of
the maximum and minimum magnitudes. Finally, one understands why the signal is
independent of $\Omega$: the coefficient $c_1$, relevant for the signal, is
determined exclusively by the relaxation {\eq{eq:dissRate}} and
isospin-to-charge conversion rates {\eq{eq:conversion}}, which do not depend
on $\Omega$. This independence of $\Omega$ is maintained even when torque
terms are then included because corrections to the solution for $\vecg{\beta}
= 0$ are of higher order in $\lambda$ and $\Omega$ and are thus disregarded in
Eqs. {\eq{eq:signal}} and {\eq{eq:signalTorque}}, except for those coming from the
torque terms along $\vecg{\lambda}$.

Comparing the two curves in {\Fig{fig:deltaI}{}}(a) we note that the impact of
the torque terms on the signal current become quite significant. Remarkably,
the isospin-torque correction to the signal current may be of the same order
as the signal current itself when entering the Coulomb blockade regime. In
{\Fig{fig:deltaI}}(b) we plot the ratio $\Delta I_{\beta \neq 0} \text{/}
\Delta I_{\beta = 0}$, which can achieve values even as large as 2 for the
parameters chosen. The reason for this may be inferred from
{\Eq{eq:signalTorque}}: when tuning away from resonance in either direction,
$| \phi' |$ and therefore $\kappa$ in {\Eq{eq:signalTorque}} quickly reaches a
maximum [see {\Fig{fig:rates}}(b)] and simultaneously either $\Gamma^0
\text{/} \Gamma^1$ or $\Gamma^1 \text{/} \Gamma^0$ rises [see
{\Fig{fig:rates}}(a)]. Beyond the maximum the latter effect dominates,
sustaining a further increase of the ratio $\Delta I_{\beta \neq 0} \text{/}
\Delta I_{\beta = 0}$.

To identify which terms in the kinetic equations {\eq{eq:kinEq}} are
responsible for this correction, we compare in {\Fig{fig:deltaI}}(c) the $z$
component of the total isospin $\vecg{\tau}$ when torque terms are included
with the $z$ component of $\vecg{\tau}^1$ for the case when we only keep the
term $\dot{\vecg{\tau}}^0 \sim - \vecg{\beta} \times \vecg{\tau}^1$ in
{\Eq{eq:kinEq}}. Clearly, this term is sufficient to reproduce the total
isospin polarization for $V_g > 0$. The above suggests that the torque-induced
isospin polarization is the result of a two-step mechanism: first, a charge
transition from $n = 1$ to $n = 0$ occurs in the SQD, accompanied by an
\emph{induced coherent precession} with frequency $\left| \vecg{\beta}
\right|$ of the isospin. After that, a dissipative transition to charge state
$n = 0$ takes place with rate $\Gamma^0$. By these two steps, the isospin
$\vecg{\tau}^1$ experiences effectively the effective magnetic field \
$\vec{B}^0 = \vec{\Omega} + \vecg{\lambda}$ [due to $\dot{\vecg{\tau}}^1 \sim
\left( \vec{\Omega} + \vecg{\lambda} \right) \times \vecg{\tau}^1$] plus an
additional, noncollinear contribution along $- \vecg{\beta}$ (due to
$\dot{\vecg{\tau}}^0 \sim - \vecg{\beta} \times \vecg{\tau}^1$). We checked
that the torque term $\dot{\vecg{\tau}}^1 \sim \vecg{\beta} \times
\vecg{\tau}^1$ in {\Eq{eq:kinEq}} is not important here by simply leaving it
out. Thus, the total effective field $\vec{B}^0 - \vecg{\beta}$ is slightly
rotated towards $+ \vecg{\lambda}$ if $\phi' > 0$ (as for positive $V_g$, cf.
{\Fig{fig:rates}}) as compared to $\vec{B}^0$. Since $\vecg{\tau}^1$ tends to
orient itself antiparallel to these effective magnetic fields, it acquires a
component along $- \vecg{\lambda}$, resulting in a {\emph{decrease}} in
$\tau_z$ [cf. {\Fig{fig:deltaI}}(c)]. A similar analysis shows that for
negative $V_g$ the dominant effect comes rather from the charge-state-conserving torque term $\dot{\vecg{\tau}}^0 = \vecg{\beta} \times
\vecg{\tau}^0$ in {\Eq{eq:kinEq}}.

\subsection{Differential readout conductance}

We next discuss the differential signal conductance
\begin{eqnarray}
  \Delta G & = & \frac{\partial \Delta I}{\partial V},  \label{eq:condsignal}
\end{eqnarray}
which is directly measured in experiments {\cite{Barthel10}}. Our findings are
summarized in {\Fig{fig:cond1}}, which compares{\footnote{By
{\Eq{eq:signal}}-{\eq{eq:signalTorque}} the $\lambda$-dependence of the signal
current comes entirely from $\Delta I_{\beta = 0}$, i.e., $\Delta I \sim
\lambda^2 \Gamma \text{/} T^2 $ due to the first factor $\sim \left| \vec{C}
\right|^2 \text{/} \Gamma^1$ in {\Eq{eq:signal}}. As a result, $\Delta G \sim
\lambda^2 \Gamma \text{/} T^3$ since the conductance changes with the bias on
the scale $T \gtrsim \Gamma$.}} $\Delta G \sim \lambda^2 \Gamma \text{/} T^3$,
plotted as a function of gate voltage when including and excluding the torque
terms, as well as their difference, the torque correction $\delta G = \Delta
G_{\beta \neq 0} - \Delta G_{\beta = 0} \sim \lambda^2 \Gamma \text{/} T^3
\kappa \sim \lambda^2 \Gamma^2 / T^4$.{\footnote{From
{\Eq{eq:signal}}-{\eq{eq:signalTorque}} we obtain $\delta G \sim \Delta
G_{\beta \neq 0} - \Delta G_{\beta = 0} \sim (\Delta I_{\beta \neq 0} - \Delta
I_{\beta = 0}) / T \sim | \vec{C} |^2 / \Gamma^1 \kappa \text{/} T \sim
\lambda^2 \Gamma \text{/} T^3$.}} Figures {\fig{fig:cond1}}(a) and (c) corroborate that the
relative impact of the torque terms on the conductance signal, $\delta G
\text{/} \Delta G \sim \kappa$, becomes larger when $\kappa \sim \Gamma
\text{/} T$ is increased. This is generally expected for renormalization
effects. {\Fig{fig:cond1}}(c) illustrates that an asymmetry of tunnel rates
$\gamma = \sqrt{\Gamma^s \text{/} \Gamma^d} > 1$ enhances the torque effects
as well. This is also expected, since in this case the SQD is emptied less
often than it is filled, leaving it nearly always singly occupied. Then the
coupled SQD-qubit undergoes long periods of \emph{coherent} time evolution
and the torque terms can precess the isospin more effectively than for the
opposite asymmetry $\gamma \leqslant 1$.

{\Figure{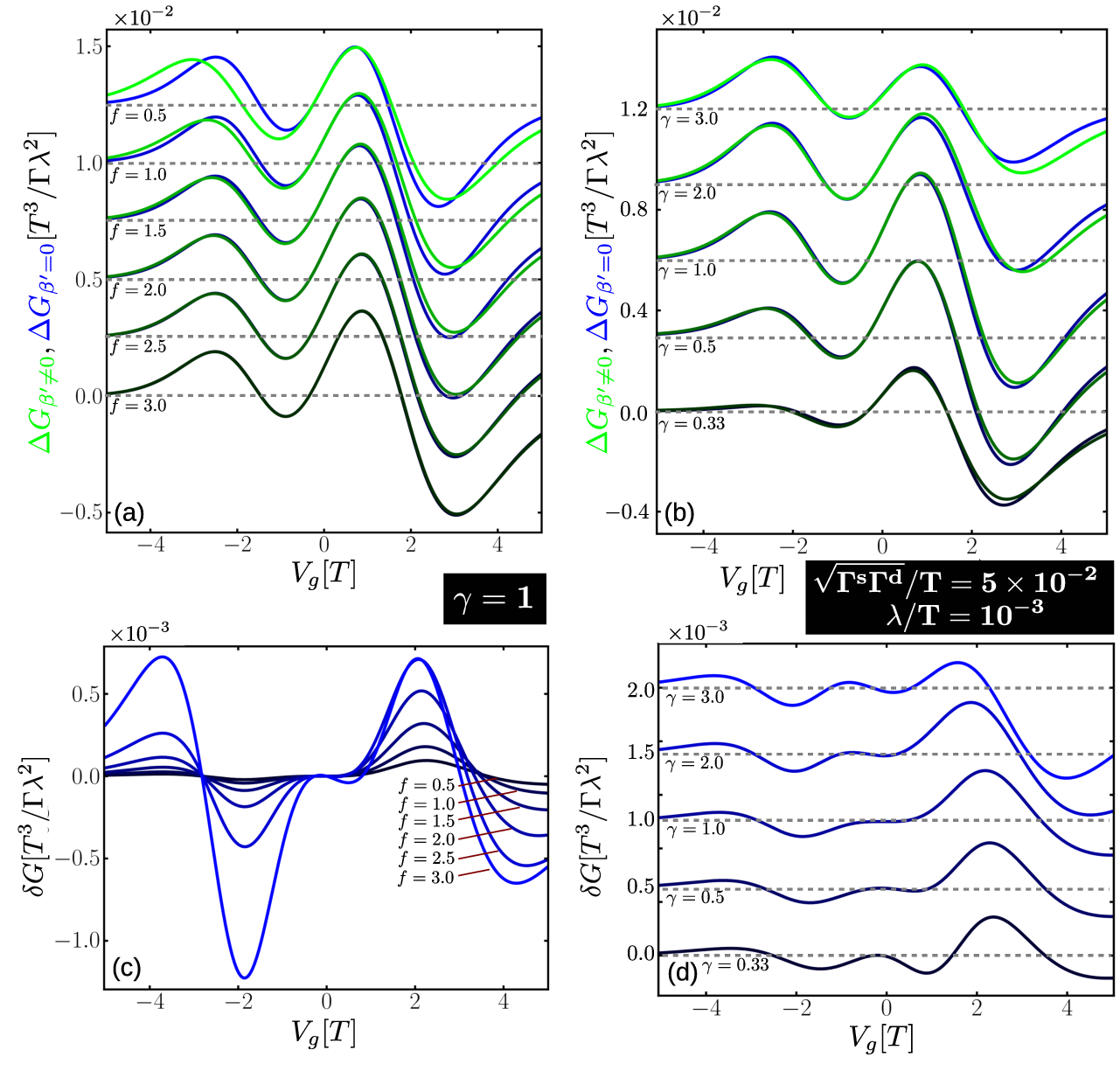}{Differential signal conductance \
{\eq{eq:condsignal}} at finite bias $V_b = 3 T$ as a function of gate voltage
$V_g = - \varepsilon$. In all plots, we set $D \text{/} T = 10^3$. $\Delta G$
is shown both including torque terms (green) and neglecting them (blue) in
(a,c) and their difference $\delta G$ is shown in (b,d). In (a,b) we compare
the results for coupling strengths, $\Gamma^s \text{/} T = \Gamma^d
\text{$\text{/}$} T = 10^{- f}$ and $\lambda \text{/} T = 10^{- 2 f}$ for
symmetric tunnel rates $\gamma = 1$, while in (b,d) we change the tunnel
couplings asymmetry $\gamma = \sqrt{\Gamma^s \text{/} \Gamma^d}$ while keeping
$\sqrt{\Gamma^s \Gamma^d} \text{/} T = 5 \cdot 10^{- 2}$ and $\lambda \text{/}
T = 10^{- 3}$ fixed. The curves in (a), (c), and (d) are vertically offset for different
parameters with the $\Delta G \text{/} \delta G = 0$ line indicated by the
gray dotted lines. The axis labels refer to the lowest curve. Note the units
in (a) and (b), which account for the scaling of the conductance signal $\Delta G
\sim \lambda^2 \Gamma \text{/} T^3$.\label{fig:cond1}}}

In {\Fig{fig:cond2}} we systematically investigate the impact on the two main
features of the $\Delta G$ traces of {\Fig{fig:cond1}}(a,b), namely the
position and its magnitude of the large dip at $V_g > 0$. We plot the absolute
\emph{correction} due to the isospin torque to the dip position
\begin{eqnarray}
  \delta V_g^{\mathrm{dip}} & = & V_{g, \beta \neq 0}^{\mathrm{dip}} -
  V^{\mathrm{dip}}_{g, \beta = 0},  \label{eq:deltaVg}
\end{eqnarray}
and a relative correction to its magnitude,
\begin{eqnarray}
  R & = & \frac{\Delta G^{\mathrm{dip}}_{\beta \neq 0} - \Delta
  G^{\mathrm{dip}}_{\beta = 0}}{\Delta G^{\mathrm{dip}}_{\beta = 0}}, 
  \label{eq:conductanceRatio}
\end{eqnarray}
as a functions of the bias voltage. In {\Fig{fig:cond2}}(a,b) we see $\delta
V_g > 0$ for all biases and parameters, i.e., the dip is shifted deeper into
the Coulomb blockade regime due to the isospin torque. As expected from the
above discussion of {\Fig{fig:cond1}}, the correction to the position
increases when $\Gamma / T$ rises as in {\Fig{fig:cond2}}(a) or the asymmetry
\ $\Gamma^s > \Gamma^d $ rises as in {\Fig{fig:cond2}}(b).

By contrast, {\Fig{fig:cond2}}(c) shows that the qualitative correction $R$ to
the magnitude depends on the parameters: for small bias, the dip is enhanced
$(R > 0)$ by the torque terms, while it is suppressed \ $(R < 0)$ in the limit
of large bias. Figure {\fig{fig:cond2}}(d) shows that this tendency is independent of
the asymmetry of the tunnel couplings. If the source tunneling barrier is more
transparent $(\Gamma^s > \Gamma^d)$, we find a nonmonotonic dependence with a
strong enhancement of the dip close to $V_b \sim 2 T$ that can reach up to
30\% for an asymmetry of $\Gamma^s \text{/} \Gamma^d = 9$, a typical
experimental value. In this case the dip position correction $\delta V_g$ in
{\Fig{fig:cond2}}(b) is also nonmonotonic. Again this corroborates the
increased relative importance of the torque terms due to long waiting times in
the SQD.

{\Figure{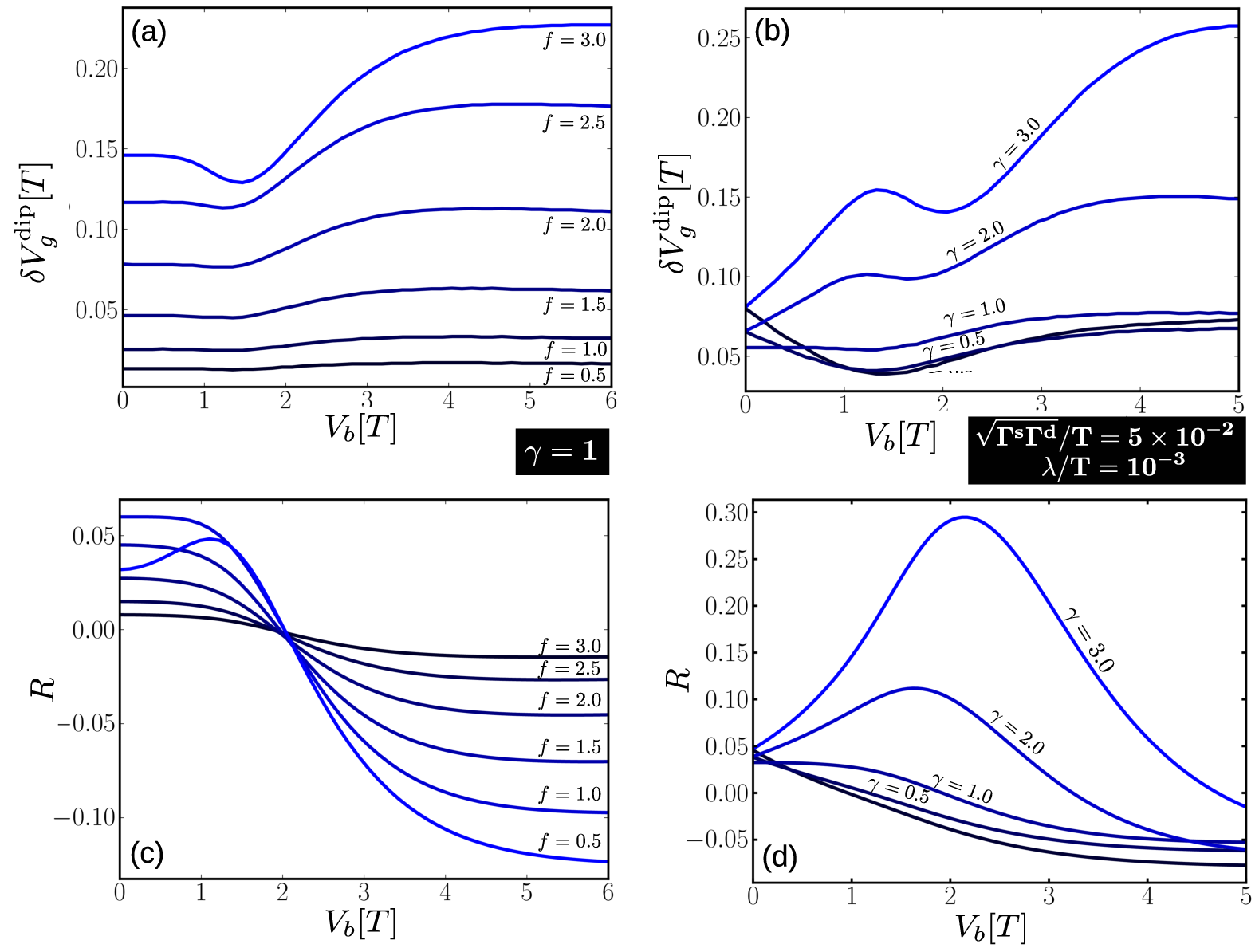}{Bias dependence of the torque
corrections. The correction $\delta V_g^{\mathrm{dip}}$, given by
{\Eq{eq:deltaVg}}, is shown in (a) for different coupling strengths $\Gamma^s
\text{/} T = \Gamma^d \text{$\text{/}$} T = \sqrt{\lambda \text{/} T} = 10^{-
f}$ and in (b) for different asymmetries $\gamma = \sqrt{\Gamma^s \text{/}
\Gamma^d} \text{}$. The relative correction $R$, given by
{\Eq{eq:conductanceRatio}}, is shown in (c) for different couplings and in (d)
for different asymmetries, chosen as in (a) and (b), respectively. All
definitions and other parameters are the same as in Figs. {\fig{fig:cond1}}(a) and
(b).{\color{blue} \label{fig:cond2}}}}

\section{Summary and Outlook}\label{sec:summary}

We have analyzed the backaction of a capacitive readout of a charge
qubit by probing the differential conductance of a nearby sensor quantum dot
(SQD). To this end we extended the kinetic equations used previously
{\cite{Gurvitz05,Gurvitz08,Shnirman98,Makhlin01a}} by including spin, local
interaction on the SQD and, most importantly, renormalization effects of (i)
the level positions of the coupled SQD-qubit system, generating qubit-isospin
torques and (ii) the tunneling rates connecting the SQD to the electrodes.
Our study, focused on the ensemble-averaged, stationary conductance
signal, already provides indications that these renormalization effects are
important for such detection schemes. In particular, at the crossover to
Coulomb blockade (the experimentally relevant regime of highest detection
sensitivity), these effects matter.

The isospin torque terms $\sim \vecg{\beta} \times \vecg{\tau}^n $ in the
kinetic equations for the coupled SQD-qubit system induce an additional
precession of charge-projected qubit isospins $\vecg{\tau}^n$. This
renormalization effect relies on the response of the SQD tunneling rate --
scaling as $\sim \Gamma$ -- to perturbations on the internal energy scales
$\sim \Omega, \lambda$ of the SQD-qubit system. This is exactly the
sensitivity that is also exploited for the readout of the qubit state. Thus,
isospin torques cannot be avoided since they incorporate terms that scale in
the same way with these parameters as the terms responsible for the readout.

We have compared these isospin torque terms with analogous terms due to the
spintronic exchange field that is found in quantum dot spin
valves {\cite{Koenig03,Martinek03b,Hauptmann08,Gaass11}}. In the latter, the
spin-dependent level renormalization that the field represents is caused by
spin-dependent tunneling rates, while the above qubit-torque terms derive from
an isospin-dependent effective level position of the electron in the SQD that
is used in the readout. A consequence of this difference in the microscopic
origin is that the isospin torque can additionally couple isospins for
different SQD charge states ($n = 0 , 1$), e.g., terms such as \
$\dot{\vecg{\tau}}^1 \sim \vecg{\beta} \times \vecg{\tau}^0$ appear, in
addition to a precession that preserves this charge state, i.e., terms of the
form $\dot{\vecg{\tau}}^n \sim \vecg{\beta} \times \vecg{\tau}^n$. The latter
are the only ones that appear in spintronics. We discussed that both types of
isospin torque terms are crucial for the description of the stationary
readout.

Furthermore, the renormalization of the SQD detector tunnel rates (level
shifts and cotunneling) is found to be crucial: without those terms, the
positivity of the density operator can be severely violated,
invalidating the approach, at least in the Markovian limit (see
{\App{sec:positivity}}). These corrections, extensively studied in transport
through QDs, have so far received little attention in the context of
quantum measurements \ and require one to go beyond the standard
Born-Markov approximation plus secular approximation. We have provided an
important, general check on any such extension by deriving a rigorous sum rule
for the charge-projected isospins that holds order-by-order in the SQD tunnel
coupling $\Gamma$. This sum rule, recently discussed in a general
setting {\cite{Salmilehto12}}, is imposed by the conservation of the qubit
isospin during tunneling and in fact holds for any qubit-SQD observable that
respects this symmetry.

The basic reason why renormalization effects are important in
weak measurements is a simple one: if an electron on the detector quantum
dot has time to probe the qubit, it also has time to fluctuate and thereby
renormalize system parameters. For the parameter regime considered here,
$\Omega, \lambda \lesssim \Gamma$, standard Born-Markov approximations,
combined with Davies' secular approximation, are not applicable, as we have
explicitly verified, and these furthermore violate the above general sum
rule.

The kinetic equations presented here provide a new starting point for studying
the impact of the isospin torque on the \emph{transient dynamics} of the
qubit Bloch vector and the {\emph{measurement dynamics}}. The
measurement-induced isospin torques lead to a modification of the relaxation
and dephasing rates of the isospin $\vec{\tau} = \vec{\tau}^0 + \vec{\tau}^1$,
which can be found by solving the kinetic equations {\eq{eq:kinEq}}
time-dependently for a given initial state. Preliminary results
indicate that the time for the exponential decay of the isospin magnitude
$\left| \vec{\tau} \right|$ to its stationary value can be
significantly altered.  However, the study of such transient
effects requires the non-Markovian corrections into {\Eq{eq:kinEq}}. Here an
interesting question is the possible additional rotation of the qubit Bloch
vector due to the torque terms during the decay. Thus, the coherent backaction
may not only be a nuisance, but could also be useful for the manipulation of
qubits due to the electric tunability of the isospin torques that we derived
here. More generally, the analogy of charge readout in quantum information
processing with spintronics of quantum dot devices may be a fruitful one to be
explored further.

\begin{acknowledgments}
  We acknowledge stimulating discussions with H. Bluhm, J. K{\"o}nig,
 L. Schreiber, J. Schulenborg, and J. Splettst{\"o}sser. We are grateful
 for support from the Alexander von Humboldt foundation.
\end{acknowledgments}

\appendix

\section{Real Time Diagrammatics}\label{sec:RTD}

We derive the kinetic equations {\eq{eq:kinEq}} for the averages occurring in
{\Eq{eq:redDens}} by applying the real-time diagrammatic technique,
{\cite{Schoeller94,Schoeller09a,Leijnse08a}} which we briefly review here to
introduce the notation and to give the starting point for the discussion of
the approximations we employ. The real-time diagrammatic technique starts from
the von-Neumann equation for the density operator of the total system,
\begin{eqnarray}
  \dot{\rho}_{\text{tot}} (t) & = & - i L \rho_{\text{tot}} (t) - i (L_T +
  L_R) \rho_{\text{tot}} (t), 
\end{eqnarray}
with the Liouvillians $L_{\alpha} \cdot = \left[ H_{\alpha}, \text{ $\cdot$ }
\right]$ for $\alpha = Q, R, S, T, I$ and $L = L_Q + L_S + L_I^{}$ mediating
the free evolution of SQD and qubit (cf. {\Sec{sec:model}}). Here, the dot ``$ \cdot $'' indicates the operator on which the superoperator $L_\alpha$ acts. Assuming a
factorizable initial state $\rho_{\text{tot}} (t_0) = \rho (t_0) \otimes
\rho^s \otimes \rho^d$, the idea is to integrate out the noninteracting leads
[cf. discussion below {\Eq{eq:reservoirs}}]. This yields a kinetic equation
for the reduced density operator $\rho = \mathrm{tr}_{\mathrm{res}}
(\rho_{\text{tot}})$:
\begin{eqnarray}
  \dot{\rho} (t) & = & - i L \rho (t) + \int_{- \infty}^{+ \infty} d t' W (t')
  \rho (t - t'),  \label{eq:redEq}
\end{eqnarray}
where the kernel $W (t')$ incorporates the effect of the leads in the past
[i.e., $W (t') = 0$ for $t' < 0$].

If the solution of {\Eq{eq:redEq}} is found, one can calculate the
time-dependent average of any qubit observable. By contrast, to compute the
average charge current $\brkt{I_r}$ from lead $r$ into the SQD, one has to
additionally calculate a current kernel $W_{I_r}$, since the current operator
$I_r = i [H_{T,} N_r]$ is a nonlocal observable. Here $N_r = \sum_{k, \sigma}
c^{\dag}_{r k \sigma} c_{r k \sigma}$ denotes the particle number operator of
lead $r$. The average current is then given by
\begin{eqnarray}
  \brkt{I_r} & = & \tr{Q + S} \int_{- \infty}^{+ \infty} d t' W_{I_r} (t')
  \rho (t - t') .  \label{eq:current}
\end{eqnarray}
Our approximations are now as follows:

(1) We first carry out a Markov approximation, i.e., we consider only changes
of the density operator $\rho (t - t')$ in the Schr{\"o}dinger picture that
take place on the time scale on which $W (t')$ decays. We therefore
approximate $\rho (t - t') \approx \rho (t)$ and express the kernel by its
Laplace transform $W (t') = \frac{1}{2 \pi} \int^{+ \infty + i 0}_{- \infty +
i 0} d z e^{- i z t'} W (z)$. Inserting this into {\Eq{eq:redEq}} yields
\begin{eqnarray}
  \dot{\rho} (t) & = & - i L_{\mathrm{eff}} \rho (t) = [- i L + W (i 0)] \rho
  (t)  \label{eq:GMEMarkov}
\end{eqnarray}
where $W (i 0) = \int_0^{\infty} d t e^{i z t} W (t) |_{z = i 0} $
is the zero-frequency component of the kernel. One can prove
{\cite{Schoeller09a,Leijnse08a}} that the stationary state calculated from
{\Eq{eq:GMEMarkov}} is the {\emph{exact}} stationary solution of
{\Eq{eq:redEq}}. Similarly, the stationary current is obtained from
{\Eq{eq:current}} by inserting the stationary density operator and replacing
the time-integrated current kernel by its zero-frequency component.

(2) We next expand the kernel in orders of the tunneling Liouvillian $L_T$ and
keep only terms up to $O (L_T^4)$. The systematic perturbative expansion of
the kernels {\eq{eq:redEq}}, {\eq{eq:current}} in powers of the tunneling
Liouvillian $L_T$ \ is derived in, e.g., {\Cite{Schoeller09a,Leijnse08a}}
together with a diagrammatic representation. The $O (L_T^{2 k})$-contribution
to the kernel schematically reads
\begin{eqnarray}
  & i W^{(2 k)} = \sum_{\mathrm{contr}} \prod_{} \gamma_i (- 1)^{N^p} & 
  \nonumber\\
  & \times G^{p_{2 k}}_{2 k} \frac{1}{i 0 + X_{2 k - 1} - L} G^{p_{2 k -
  1}}_{2 k - 1} \ldots \frac{1}{i 0 + X_1 - L} G^{p_1}_1 ; &  \label{eq:W2k1}
\end{eqnarray}
see {\Cite{Leijnse08a}} for notation and discussion. To see when higher-order
corrections in $L_T$ are important, we divide all bath frequencies in the
integrals by temperature $T$, that is, we substitute by dimensionless $x_i =
(\omega_i - \mu_i) \text{/} T$. This yields schematically
\begin{eqnarray}
  i \frac{W^{(2 k)}}{T} & = & (- 1)^{N^p} \sum_{\mathrm{contr}} \left( \prod_i
  \frac{\Gamma_i}{T} \right) I^{(2 k)} \left( \frac{L - \mu}{T} \right) 
  \label{eq:W2k}
\end{eqnarray}
\begin{eqnarray}
  & I^{(2 k)} \left( \frac{L - \mu}{T} \right) = \left( \prod_{} f_i \right)
  \times &  \label{eq:Ifun}\\
  & G^{p_{2 k}}_{2 k} \frac{1}{i 0 + \tfrac{X_{2 k - 1} - (L - \mu_{2 k -
  1})}{T}} G^{p_{2 k - 1}}_{2 k - 1} \ldots G_2^{p_2} \frac{1}{i 0 +
  \tfrac{X_1 - (L - \mu_1)}{T}} G_1 &  \nonumber
\end{eqnarray}
where $f_i$ denotes the Fermi functions. We see that $W / T$ scales as
$(\Gamma \text{/} T)^k$ multiplied with a function whose relevant energy
scales are set by $(L - \mu) \text{/} T$, i.e., the distance of the energy
difference of the reduced system to the electro-chemical potentials compared
to temperature. If $\Gamma \text{/T}$ is small, one can neglect higher-order
terms unless $I^{(2)}$ is exponentially suppressed by the Fermi functions in
Coulomb blockade. Then at least $O ((\Gamma \text{/} T)^2)$ must be included.

(3) Since we focus here on the limit of small $\lambda, \Omega$ we perform an
expansion of $W$ not only in $L_T$, but we also expand the propagators in
{\Eq{eq:Ifun}} in the Liouvillian of the qubit together with its interaction
with the SQD $L_{Q I} \assign L_Q + L_I$:
\begin{eqnarray}
  &  & \frac{1}{i 0 + x_n - (L - \mu_n) \text{/} T} \approx \nonumber\\
  &  & \left( 1 - \frac{L_{Q I}}{T} \frac{\partial}{\partial x_n} + \ldots
  \right) \frac{1}{i 0 + x_n - (L_S - \mu_n) \text{/} T} \hspace{1em} 
\end{eqnarray}
employing $[L_{Q I}, L_S] = 0$. Truncating this expansion after the first
order in $L_{Q I}$ is therefore justified if $\Omega, \lambda \ll T$. To sum
up, our approximations are valid if
\begin{eqnarray}
  & \lambda, \Omega \tilde{<} \Gamma \ll T. & 
\end{eqnarray}
In this case, we will only keep terms in $O (\Gamma, \Gamma \lambda \text{/}
T, \Omega \Gamma \text{/} T, \Gamma^2 \text{/} T)$, but we will neglect
remaining terms of higher orders in $\Gamma$, $\lambda$, and $\Omega$.

\section{Cotunneling and Positivity\label{sec:positivity}}

We emphasized in the main part that a consistent treatment can only account
for the readout (back)action terms $| \vec{C}_r | \sim \Gamma \lambda \text{/}
T$ if level renormalization effects $| \vecg{\beta}_r | \sim \Gamma \lambda
\text{/} T, \Gamma \Omega \text{/} T$ (the isospin-torque terms) are also
included. For continuous measurements $\lambda \ll \Gamma$, this in turn
requires the inclusion of the renormalization of the tunneling rates of the
SQD $\sim \Gamma^2 \text{/} T$ in {\Eq{eq:dissRate}} (see also
{\App{sec:cot}}) into the kernel $W$ (see {\App{sec:RTD}}). In this appendix
we show that aside from the consistency of the perturbation theory, an additional,
compelling reason for this is that an initially valid reduced density matrix
$\rho (0)$ can become severely nonpositive when subject to the time evolution
described by the generalized master equation {\eq{eq:GMEMarkov}}, i.e., the
dynamical linear map on density operators generated by $- i L_{\mathrm{eff}}
\assign - i L + W (i 0)$ is not positive.{\footnote{Interestingly, positivity
of $\rho (t)$ is no issue if the SQD and the qubit are decoupled: one can
recast $- i L_{\mathrm{eff}}$ for the isolated SQD given either up to $O
(\Gamma)$ or $O (\Gamma)^2$ into Lindblad form. This even rigorously proves
\emph{complete} positivity of the time evolution superoperator for this
case. On the contrary, the generator $- i L_{\mathrm{eff}}$ does not have the
Lindblad form when the coupling is finite.}} Equivalently, the solution $\rho
(t)$ can only remain a positive operator for all times $t > t_0$ if we demand
that all eigenvalues of the superoperator $L_{\mathrm{eff}}$ have nonpositive
imaginary parts (assuming for simplicity that $L_{\mathrm{eff}}$ can be
diagonalized). It is important to address this point, even though
here we are only interested in the long-time limit, i.e., the stationary
solution of {\Eq{eq:redEq}}. Non-Markovian corrections to our approximation
$L_{\mathrm{eff}} = - i L + W (i 0)$ only affect the transient approach to the
stationary state. However, if the imaginary part of an eigenvalue of
$L_{\mathrm{eff}}$ crosses zero, the degeneracy with the stationary state gives
rise to an unphysical stationary state (e.g., negative occupation
probabilities) and our approach breaks down.

The effective Liouvillian is positive up to $O (\Gamma)$ only if the torque
terms are neglected. However, this a physically inconsistent treatment since
there is no reason keep terms $\sim \vec{C}$ of order $\Gamma \lambda / T$
while neglecting terms $\sim \vecg{\beta}$ of the same order [cf. main text,
in particular {\Fig{fig:rates}}]. In {\Fig{fig:eigenvalues}}(a) we explicitly
illustrate this, plotting the largest imaginary parts of the $O (\Gamma)$
effective Liouvillian as function of the gate voltage at fixed finite bias,
both excluding the torque terms (dashed lines) and including them (solid
lines). The positivity violation starts at the crossover into the Coulomb
blockade regime and then persists. The violation can be traced to the
charge-state mixing isospin-torque terms [see {\Eq{eq:kinEq}}].

Figure {\fig{fig:eigenvalues}}(b) reveals that if contributions $O (\Gamma^2)$ are
consistently included in $L_{\mathrm{eff}}$, no exponentially increasing modes
occur well into the Coulomb blockade regime. This is another indication that
$O (\Gamma^2_{})$-contributions inevitably {\emph{must}} be accounted for when
level renormalization effects are considered. It is interesting to note
here that a standard method to enforce the positivity of the reduced density
matrix when deriving kinetic equations is the {\emph{secular approximation}}.
{\cite{Breuer,Chirolli08}} However, in {\App{sec:sumRule}}, we explain that
this approximation is not applicable and moreover does not comply with an
exact isospin sum rule (which expresses a conservation law), whereas our
treatment does.

Finally, for completeness we indicate why the solution of {\Eq{eq:GMEMarkov}}
stays positive for all times only if the nonzero eigenvalues have positive
imaginary parts, following a reasoning similar to {\Cite{Terhal00}}. We assume
that the effective Liouvillian can be diagonalized:
\begin{eqnarray}
  L_{\mathrm{eff}} (i 0) \bullet & = & \sum_i \alpha_i A_i  \Tr
  (\tilde{A}_i^{\dag} \bullet) 
\end{eqnarray}
where $\bullet$ denotes the operator that the effective Liouvillian is applied
to. Since we consider the zero-frequency effective Liouvillian, all
eigenvalues $x_i$ are either purely imaginary or they appear in pairs
$\alpha_i^{\pm} = \pm a_i + i b_i$. The corresponding right and left
eigenoperators $A_i^{\pm}$ and $\tilde{A}_i^{\pm}$, respectively, are then
Hermitian in the first case or they come in Hermitian conjugate pairs in the
second case. This property ensures that the solution of Eq.
(\ref{eq:GMEMarkov}) stays Hermitian while the probability conservation
follows from the fact that the operators $A_i$ are either traceless or
correspond to eigenvalue with $a_i = 0$.{\cite{Schoeller09a}} The right
eigenvector of the zero eigenvalue, assumed to be unique and labeled by $i =
0$, is associated with the stationary state, i.e., $A_0 = \rho_{\infty}$ with
$\mathrm{tr} (\rho_{\infty}) = 1$ and $\tilde{A}_0 = \mathbbm{1}$ (this follows
from probability conservation). The formal solution of {\Eq{eq:GMEMarkov}}
thus reads
\begin{eqnarray}
  \rho (t) & = & e^{- i L_{\mathrm{eff}} t} \rho (t_0) \nonumber\\
  & = & \rho_{\infty} + \sum_{\substacktwo{i > 0}{\eta = \pm}} e^{(b_i - \eta
  i a_i) t} A_i^{\eta}  \Tr (\tilde{A}^{\eta}_i \rho (t_0)) .  \label{eq:sol}
\end{eqnarray}
Hermiticity and probability conservation guarantee that the eigenvalues of
operator (\ref{eq:sol}) are real and sum up to 1. However, this does not
exclude negative eigenvalues, in which case the positivity of $\rho (t)$ is
violated. This will happen if at least one exponentially increasing mode with
$b_i > 0$ contributes to {\eq{eq:sol}}, because $\rho (t)$ is unbounded in
that case for $t \rightarrow \infty$. Thus, some of its eigenvalues will also
be unbounded and one of those has to be negative if the sum of all eigenvalues
is fixed to 1.

{\Figure{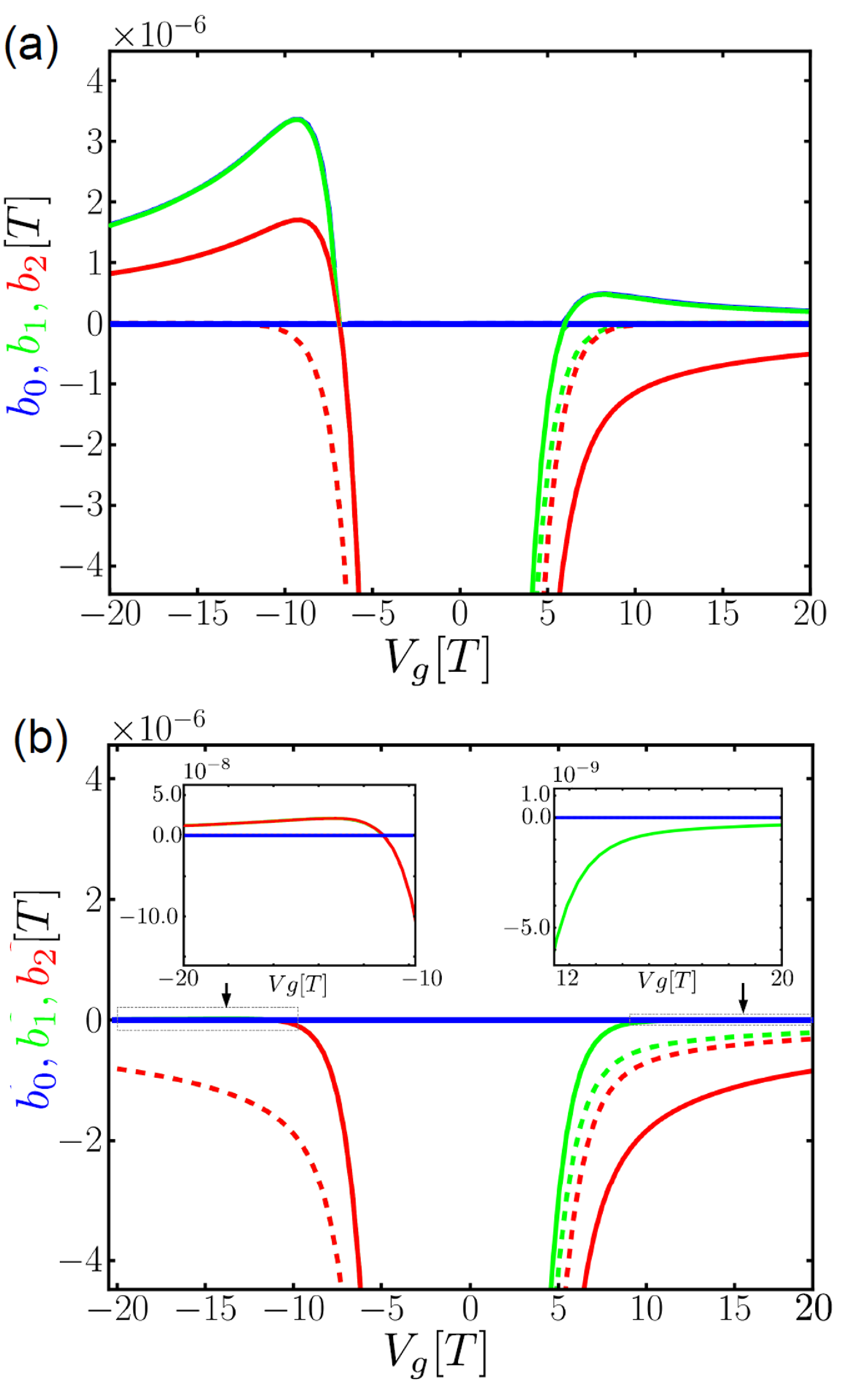}{Comparison of the first three largest
imaginary parts $b_0$, $b_1$, and $b_2$ of the eigenvalues of $L_{\mathrm{eff}}$
plotted as a function of gate voltage. The perturbation expansion of the
kernel $W$ in (a) is taken up to $O (\Gamma)$ and in (b) up to $O (\Gamma^2)$.
The eigenvalues are shown both including torque terms $\sim \vecg{\beta}$
(bold lines) and neglecting them (dashed lines). The remaining parameters are
$V_b \text{/} T = 5$, $\Gamma_L \text{/} T = \Gamma_R \text{/} T = 10^{- 1}$,
$\Omega \text{/} T = \lambda \text{/} T = 10^{- 2}$ \ and $D \text{/} T =
10^3$. Notably, the naive $O (\Gamma)$-approximation for $W$, i.e., including
the torques but not the renormalization of the SQD rates, inevitably leads to
eigensolutions exponentially increasing with time, cf. {\Eq{eq:sol}}. This is
due to eigenvalues with positive imaginary part that appear in (a) for $| V_g
| \gtrsim V_b$, i.e., when entering the Coulomb blockade regime. The
consistent inclusion of $O (\Gamma^2)$ terms in (b) prevents the occurrence of
exponentially increasing modes. Only for large negative gate voltages,
slightly positive eigenvalues exist, taking only a fraction of 1/100 of the
value compared to the $O (\Gamma)$ case in (a). We attribute these to a
neglect of corrections of even higher orders in $\Gamma, \lambda$ and
$\Omega$.\label{fig:eigenvalues}}}

\section{Renormalized SET rates\label{sec:cot}}

The renormalized SQD rates, {\Eq{eq:dissRate}}, may be rewritten as
\begin{eqnarray}
  \Gamma^{0, 1}_r (\varepsilon) & = & \Gamma_r \left( f_r^{\pm} (\varepsilon)
  \pm (f^+_r (\varepsilon))' \sum_{q = s, d} \tfrac{\Gamma_q}{2 T} \phi_q
  (\varepsilon) \right) \nonumber\\
  &  & \mp \sum_{q = s, d} \tfrac{\Gamma_r \Gamma_q}{2 T} \phi'_r
  (\varepsilon) (2 f^+_q (\varepsilon) + f^-_q (\varepsilon)) \\
  & \approx & \Gamma_r f_r^{\pm} \left( \varepsilon + \sum_{q = s, d}
  \tfrac{\Gamma_q}{2 T} \phi_q (\varepsilon) \right) \nonumber\\
  &  & \mp \sum_{q = s, d} \tfrac{\Gamma_r \Gamma_q}{2 T} \phi'_r
  (\varepsilon) (f^+_q (\varepsilon) + 1) . 
\end{eqnarray}
In the first term we used $f^{\pm}_r (\omega) = f (\pm (\omega - \mu_r)
\text{/} T)$ and therefore $(f^+_r)' = \pm (f^{\pm}_r)'$, whereas in the
second term $f^+_q + f^-_q = 1$. Clearly, the first correction term
corresponds to the change in the tunneling rates be virtual fluctuations that
effectively shift the level position to $\varepsilon' = \varepsilon + \sum_q
\tfrac{\Gamma_q}{2 T} \phi_q (\varepsilon)$. The dependence of the second term
on the level position is typical of elastic cotunneling. Even when $f^+_q
(\varepsilon)$ is exponentially small for $\varepsilon - \mu_r \gg T$, the
term $\sum_q \tfrac{\Gamma_r \Gamma_q}{2 T} \phi'_q (\varepsilon) \approx
\sum_q \tfrac{\Gamma_q \Gamma_r}{2 \pi (\varepsilon - \mu_r)}$ only decays
algebraically. This yields a finite, positive relaxation rate $\Gamma^1_r$ for
charge state 1, which ensures the positivity of the density matrix.

\section{Sum Rules and Conservation Laws}\label{sec:sumRule}

In this Appendix, we derive and generalize the sum rule {\eq{eq:tausum}} for
the isospins in the main text. We start by noting that the total isospin
operator, $\op{\vecg{\tau}} = \sum_n \op{\vecg{\tau}}^n$, only acts on the
qubit part, in contrast to the charge-projected ones, $\op{\vecg{\tau}}^n =
\op{P}^n  \op{\vecg{\tau}}$. Exploiting {\Eq{eq:redEq}}, the time evolution of
its average $\vecg{\tau} (t) = \Tr_{Q + S}  \op{\vecg{\tau}} \rho (t)$ is
given by
\begin{eqnarray}
  \dot{\vecg{\tau}} (t) & = & \tr{Q}  \op{ \vecg{\tau}} \int_0^{\infty} d t' 
  \tr{S} [- i L \delta (t') + W (t')] \rho (t - t') \nonumber\\
  & = & - i \tr{Q + S} \op{\vecg{\tau}} L \rho (t)  \label{eq:sumRuleSpin}
\end{eqnarray}
where the kernel-induced part vanishes because \ the kernel satisfies the sum
rule $ \Tr_S W (t') = 0$ that guarantees probability
conservation.{\cite{Schoeller09a,Leijnse08a}} This statement holds
individually for contributions to $W$ of each order in $\Gamma$. The
right-hand side of {\Eq{eq:sumRuleSpin}} now follows from Eqs.
{\eq{eq:tau0free}}-{\eq{eq:tau1free}} and gives the sum rule {\Eq{eq:tausum}}
of the main text. In physical terms, it expresses the fact that the
isospin is conserved by the tunneling, i.e. $\left[ H_T, \op{\vecg{\tau}}
\right] = 0$. Such constraints on kinetic equations have recently been
investigated on a general level in {\Cite{Salmilehto12}}, where a generalized
{\emph{current}} conservation law is set up. It expresses the idea that the
time evolution of a reduced system observable $\op{A}$ can only be correctly
reproduced by a generalized master equation if the change in this observable
induced by the kernel{\footnote{The kernel is related to the ``generalized
dissipator'' introduced in {\Cite{Salmilehto12}}.}} equals the change induced
by the system-environment interaction. For our formulation of the kinetic
equation this requirement reads \ $\Tr_{Q + S}  \op{A}  \int_0^{\infty} d t' W
(t') \rho (t - t') = - i \underset{}{\mathrm{Tr}} \left[ H_T, \op{A} \right]
\rho^{\mathrm{tot}} (t)$. If we insert the isospin $\op{\vecg{\tau}}$ for
$\op{A}$, the right-hand side is zero, which yields the isospin sum rule if
the free qubit evolution is added. The authors of {\Cite{Salmilehto12}} point
out that this is not guaranteed by all approaches used to derive kinetic
equations, in particular when a secular approximation is employed, c. f.
{\App{sec:secular}}. Our sum rule therefore provides an important consistency
check, which is clearly fulfilled by our kinetic equations (\ref{eq:kinEq}).

To show more generally that real-time diagrammatics respects internal
conservation laws of the reduced system, we next consider the more general
case of any observable $\op{A}$ that is conserved in the tunneling: $[\op{A},
H_T] = 0$. The time derivative of its average reads as
\begin{eqnarray}
  \dot{A} (t) & = & \underset{Q + S}{\Tr}  \op{A} \int_0^{\infty} d t'  [- i L
  \delta (t') + W (t')] \rho (t - t') \nonumber\\
  & = & - i \underset{Q}{\Tr} A L \rho (t) .  \label{eq:sumRule}
\end{eqnarray}
We emphasize that {\Eq{eq:sumRule}} by no means implies that our model
describes a backaction-evading / quantum nondemolition measurement of $A$: the
statistics of $A$ still changes due to the tunneling-induced change in the
reduced density matrix. The operator $\op{A}$ is still subject to the free
evolution and it is therefore not a constant of motion (as a Heisenberg
operator), which would be sufficient for a QND measurement.

The proof of {\Eq{eq:sumRuleSpin}} is particularly simple because the electron
reservoir only couples to the SQD part $\mathcal{H}_S$, while all the qubit
observables only act on $\mathcal{H}_Q$ of the Hilbert space of the reduced
system $\mathcal{H}_{\mathrm{red}} = \mathcal{H}_S \otimes \mathcal{H}_Q$.
Equation {\eq{eq:sumRule}} follows from the general observation that the kernel is a
reservoir trace of a commutator with $H_T$: $W \bullet = \Tr_{\mathrm{res}} L_T
X \bullet = \Tr_R  [H_T, X \bullet]$ where $X$ is some superoperator
expression that is irrelevant here. Then the second term in {\Eq{eq:sumRule}}
vanishes by cyclic invariance of the total trace:
\begin{eqnarray*}
  \tr{Q + S}  \op{A} W (t') \bullet & = & \tr{Q + S}  \op{A}  \tr{\mathrm{res}} 
  [H_T, X \bullet]\\
  & = & \tr{Q + S}  \tr{\mathrm{res}}  \left[ H_T, \op{A} X \bullet \right] =
  0.
\end{eqnarray*}
This general structure of the kernel $W$ is most easily seen in the
Nakajima-Zwanzig formulation, equivalent to the real-time approach used here
{\cite{Timm08,Koller10}}. To recover this structure from the diagrammatic
rules first for the zero-frequency kernel, we re-express the
contraction function $\gamma_{2 k, j} = \Tr_{\mathrm{res}} (J_{2 k}^{p_{2 k}}
J_j^{p_j}) $ in {\Eq{eq:W2k1}} involving the left-most vertex with label $2
k$, shift $J_{2 k}^{p_{2 k}}$ to its original position next to $G^{p_{2 k}}_{2
k}$, and sum over the indices associated with $2 k$. This restores the
tunneling Liouvillian and {\Eq{eq:W2k}} then schematically reads as
\begin{eqnarray}
  W^{(2 k)} & \sim & \tr{\mathrm{res}}  (L_T J_{2 k}^{p_{2 k}})  \frac{1}{i 0 +
  X_{2 k - 1} - L} \ldots G_1^{p_1} . 
\end{eqnarray}
This proof can be worked out analogously for the kernel in time
representation without a Markov approximation as it occurs in
{\Eq{eq:sumRule}}, since it has a similar structure with the propagator
denominators replaced by exponentials of the form $e^{- i (X_{2 k - 1} - L)
\Delta t}$ and integrations over all time differences (see Refs.~\onlinecite{Saptsov12a} and~\onlinecite{Reckermann13}). 

\section{Secular approximation and sum rule}\label{sec:secular}

A common procedure to avoid positivity problems arising when deriving
generalized master equations, e.g., from a Born-Markov
approximation,{\cite{Breuer,Chirolli08}} is to perform a secular
approximation. Here one decouples the occupations and secular coherences of
the eigenstates (i.e., states with the same energy) of the reduced system from
their nonsecular coherences (i.e., states with different energies). Following
the procedure described in {\Cite{Breuer}} and transforming back to the
Schr{\"o}dinger picture, we obtain the following generalized master equations:
\begin{eqnarray}
  \dot{p}^0 & = & - \dot{p}^1,  \label{eq:p0Sec}\\
  \dot{p}^1 & = & + 2 \Gamma^+ p^0 - \Gamma^- p^1 + \vec{C} \cdot E^1 \cdot
  \vecg{\tau}^1,  \label{eq:p1Sec}\\
  \dot{\vecg{\tau}}^0 & = & - 2 \Gamma^+ \vecg{\tau}^0 + \Gamma^- E^0 \cdot
  E^1 \cdot \vecg{\tau}^1 + \vec{\Omega} \times \vecg{\tau}^0, 
  \label{eq:t0ec}\\
  \dot{\vecg{\tau}}^1 & = & + 2 \Gamma^+ E^1 \cdot E^0 \cdot \vecg{\tau}^0 -
  \Gamma^- \vecg{\tau}^1 - \left( E^1 \cdot \vec{C} \right) (2 p^0 + p^1)
  \nonumber\\
  &  & + \left( \vec{\Omega} + \vecg{\lambda} - E^1 \cdot \vec{\beta} \right)
  \times \vecg{\tau}^1 .  \label{eq:t1Sec}
\end{eqnarray}
The above equations are different from our result {\eq{eq:kinEq}} in three
different respects: First, due to the Born approximation, they involve only
the leading-order tunneling rates $\Gamma^{\pm} = \sum_r \Gamma_r f_r^{\pm}
(\varepsilon)$. Second, when the charge state of the SQD is changed, the
isospins are projected by $E^n = \vec{e}^n \left( \vec{e}^n \right)^T$ onto
the directions of charge-dependent effective magnetic fields $\vec{e}^0 =
\vec{\Omega} \text{/} \Omega$ and $\vec{e}^1 = \left( \vec{\Omega} +
\vecg{\lambda} \right) \text{/} \sqrt{\Omega^2 + \lambda^2}$. Thus, the
occupations only couple to $\vecg{\tau}^1$ because $\vec{e}^0$ and $\vec{C}
\propto \vecg{\lambda}$ are orthogonal and only $\vec{e}^1$ and $\vec{C}$ have
a finite scalar product. This is a consequence of the secular approximation,
which also suppresses the tunneling-induced torque terms that couple different
charge states in our equations {\eq{eq:kinEq}} (although they have a strong
impact, cf. the last two paragraphs in {\Sec{sec:readout}}). Third, since the
Markov approximation in {\Cite{Breuer}} is carried out in the interaction
picture, the effective magnetic fields acting on the isospin within each
charge state are also different.

The stationary solution of the above kinetic equations
{\eq{eq:p0Sec}}-{\eq{eq:t1Sec}} is identical to that obtained when we neglect
cotunneling corrections and the tunneling-induced torque terms in our kinetic
equations {\eq{eq:kinEq}}. To understand this, recall that the stationary
charge-dependent isospins $\vecg{\tau}^n$ are pointing in the direction of the
effective magnetic fields acting in the respective charge state, i.e.,
$\vecg{\tau}^n = \tau^n \vecg{e}^n$ (cf. {\Eq{eq:ansatz}} and the related
discussion there). This can also be found for the stationary solution of Eqs.
{\eq{eq:p0Sec}}-{\eq{eq:t1Sec}}. Inserting the equivalent statement
$\vecg{\tau}^n = E^n \vecg{\tau}^n$ into our kinetic equations
{\eq{eq:kinEq}}, one can readily obtain \ Eqs.
{\eq{eq:p0Sec}}-{\eq{eq:t1Sec}}.{\footnote{The Lamb shift term in
{\Eq{eq:t1Sec}} is irrelevant for the stationary solution of $\vecg{\tau}^1$
because it satisfies $\vec{e}^1 \times \vecg{\tau}^1 = 0$.}}

It has in fact been shown by Davies {\cite{Davies74,Davies75}} that the
Born-Markov plus a secular approximation become exact when approaching the
limit of zero coupling (here the tunneling rate $\Gamma$) between ``system''
(qubit plus SQD) and ``environment'' (the leads) for large times $t
\rightarrow \infty$.{\footnote{In {\Cite{Davies74}}, the time is rescaled as
$\tau = g^2 t$ where $\tau$ is finite and the coupling $g \rightarrow 0$, cf.
remarks in {\Cite{Fleming10}}.}} In the limit of weak tunnel coupling $\Gamma
\ll \lambda, \Omega$, we have checked that the torque-induced corrections to
the stationary conductance become negligible. To approach this limit, we only
performed a leading-order expansion of the kernel in $\Gamma$, but not in
$\lambda $, $\Omega$ as described in {\App{sec:RTD}}. Thus, our
results comply with a Born-Markov plus secular approximation. However, we
consider a completely different situation in this paper, namely that of a weak
measurement for which the tunnel coupling $\Gamma$ is much {\emph{larger}}
than the internal energy scales $\Omega$ and $\lambda$ of the ``system''.
Thus, the occupations and the coherences of the density matrix within one
charge state of the SQD do not decouple and a Born-Markov plus secular
approximation is simply not valid in this parameter regime.

Finally, we mention that Eqs. {\eq{eq:p0Sec}}-{\eq{eq:t1Sec}} violate the
isospin sum rule {\eq{eq:tausum}}: The time derivative of the total isospin
reads as [cf. {\Eq{eq:sumRuleSpin}}]:

\begin{eqnarray}
  \dot{\vecg{\tau}}^0 + \dot{\vecg{\tau}}^1 & = & \dot{\vecg{\tau}}
  |_{\mathrm{int}} \nonumber \\ 
  &  & + 2 \Gamma^+ (E^1 \cdot E^0 -  \mathbbm{1} ) \cdot \vecg{\tau}^0
   \nonumber \\
  &  & + \Gamma^- (E^0 \cdot E^1 -   \mathbbm{1} ) \cdot \vecg{\tau}^1
   \nonumber \\
  &  & - \left( E^1 \cdot \vec{C} \right) (2 p^0 + p^1) \nonumber \\
  &  & - \left( E^1 \cdot \vecg{\beta} \right) \times \vecg{\tau}^1.
\end{eqnarray}
In the stationary limit, it is trivially fulfilled,
but for time-dependent solutions, this may not be the case. Thus, our
model provides a physically relevant example illustrating the importance of
the findings of {\Cite{Salmilehto12}} to qubit measurements: The step in the
derivation of Eqs. {\eq{eq:p0Sec}}-{\eq{eq:t1Sec}} that leads to a violation
of the sum rule (expressing the violation of the current conservation
discussed in {\Cite{Salmilehto12}}) is precisely the secular approximation.
Further studies {\cite{Celio89,Fleming10}} discussing different systems have
also indicated that the secular approximation can give rise to strong
deviations of the solution of the secular-approximated equations compared to
that obtained by more accurate approximations. This does not contradict the
results in Refs.{~\onlinecite{Davies74,Davies75}} as the proof only considers the large-time
limit. To sum up, care has to be taken, when the secular approximation is
invoked; it may capture the physics incorrectly.

\bibliographystyle{apsrev}
\end{document}